\documentclass[runningheads]{lncse}

\usepackage{amsmath}
\usepackage{amsfonts}
\usepackage{graphicx}
\usepackage{amssymb}

\begin{document}

\title{Time Step Expansions and the Invariant Manifold Approach to Lattice
Boltzmann Models}
\titlerunning{The Invariant Manifold Approach to LBMs}
\author{David J. Packwood \and Jeremy Levesley  \and  Alexander N. Gorban }
%\index[aut]{Packwood@David J. Packwood}
%\index[aut]{Levesley@Jeremy Levesley}
%\index[aut]{Gorban@Alexander N. Gorban}
\authorrunning{ D.J. Packwood \emph{et al}}
%\institute{Department of Mathematics, University of Leicester,
%LE2 2PQ, United Kingdom}
\institute{Department of Mathematics, University of Leicester,
LE2 2PQ, United Kingdom}

%\newtheorem{axiom}{Axiom}
%\newtheorem{hypothesis}{Hypothesis}
%
%\usepackage{amsmath}
%\usepackage{amsfonts}
%\usepackage{graphicx}
%
%%\oddsidemargin -0.5in
%%\textwidth 7.5in 
%
%\begin{document}

\bibliographystyle{plain}

\maketitle
\begin{abstract}
The classical method for deriving the macroscopic dynamics of a \index{Lattice Boltzmann models}lattice Boltzmann system is to use a combination of different approximations and expansions. Usually a \index{Chapman-Enskog analysis}Chapman-Enskog analysis is performed, either on the continuous Boltzmann system, or its discrete velocity counterpart. Separately a discrete time approximation is introduced to the discrete velocity Boltzmann system, to achieve a practically useful approximation to the continuous system, for use in computation. Thereafter, with some additional arguments, the dynamics of the Chapman-Enskog expansion are linked to the discrete time system to produce the dynamics of the completely discrete scheme.
\par In this paper we put forward a different route to the macroscopic dynamics. We begin with the system discrete in both velocity space and time. We hypothesize that the alternating steps of advection and relaxation, common to all lattice Boltzmann schemes, give rise to a \index{Slow manifold}\emph{slow invariant manifold}. We perform a time step expansion of the discrete time dynamics using the invariance of the manifold. Finally we calculate the dynamics arising from this system. 
\par By choosing the fully discrete scheme as a starting point we avoid mixing approximations and arrive at a general form of the microscopic dynamics up to the second order in the time step. We calculate the macroscopic dynamics of two commonly used lattice schemes up to the first order, and hence find the precise form of the deviation from the Navier-Stokes equations in the dissipative term, arising from the discretization of velocity space.
\par Finally we perform a short wave perturbation on the dynamics of these example systems, to find the necessary conditions for their stability.

\end{abstract}

\section{Introduction}
The Boltzmann Equation is a key tool within statistical mechanics, used to describe the time evolution of gases, and with some extensions, other fluids also. In this work we are concerned with calculating the dynamics of an ideal gas. This is
achieved by calculating the statistical behaviour of single particles, that is the distribution of their positions in phase space.
In particular by fixing a time point and integrating across velocity space, it is possible to calculate the macroscopic quantities
of a fluid across space. Performing such an integration allows us to take a 'snapshot' of the dynamics at any time point.
If we are concerned, however, with discovering the rates at which the macroscopic variables change, we need to apply
additional techniques and assumptions to the \index{Boltzmann equation}Boltzmann Equation. A common choice of technique to derive these macroscopic dynamics is
the 'Chapman-Enskog' procedure \cite{chapmancowling,succitlbe}. This method involves calculating the dynamics of the distribution of the particles at different orders, within
the Boltzmann Equation following a perturbation by a small parameter, the Knudsen number. Under such a perturbation
the convective dynamics of the fluid will appear at the zero order, and the dissipative dynamics at the first order. The final
result is that following such a treatment, the \index{Navier-Stokes equations}Navier-Stokes equations will be revealed. At the second-order and beyond additional terms give rise to Burnett and super-Burnet type equations repectively. 

\par Despite the Boltzmann Equation recovering the Navier-Stokes equations, to first order in the Knudsen number at least, a practical investigation into the macroscopic
dynamics of the Boltzmann system is not neccesarily complete. In order to solve the Boltzmann equation numerically,
a discretization of both time and velocity space is necessary. To be clear, in many cases  for lattice-Boltzmann methods, a discretization of space is not
necessary, as a good choice of velocity set and time step size can result in the Boltzmann equation being solved on a discrete
lattice subgroup of space points, hence suffering no extra error in this regard. 
\par Within the discrete velocity, continuous time scheme the method of choice to evaluate the macroscopic dynamics of the system has remained to be the Chapman-Enskog procedure. This is presented in a number of different ways \cite{karlince,luoce}, however the crux of the approach remains the expansion in the small parameter the Knudsen number. 
%the statistical ratio between the time particles spend under free flight and colliding, or alternatively the \emph{rate} at which particles, on average, collide per unit time. Of course in continuous time this is exactly the ratio between convective (free flight) and diffusive (collision) dynamics, making it an excellent choice of the parameter for an order by order expansion.  
The necessary discretization of the velocity performed and described in some detail in Sec. \ref{sec:discrete}, already gives rise to an additional error from the Navier-Stokes equations. Due to the approximation to the \index{Maxwell distribution}Maxwell distribution, the Navier-Stokes equations are only recovered exactly as the Mach number tends to zero. Where the Mach number is large, additional viscous (dissipative) terms appear\cite{bulkandshear}
\par In order to move from continuous time to the lattice system, an Euler step is used to approximate the time derivative. Collisions should happen exactly once per time step, therefore the rate at which collisions occur is given by the time step itself. Furthermore, if the time step is small, in the same scale as the Knudsen number inherited from the continuous time system, the asymptotics of the order by order expansion may be compromised. Bearing in mind the complexity of combining the necessary number of terms from the Euler approximation, along with the existing expansions from the continuous system and taking into account the additional small parameter, we present, in this work, a possibly simpler route to the macroscopic dynamics.
\par Our starting point is, in fact, not the continuous Boltzmann equation, but the discrete time system itself, in this sense we work in parallel to the historic lattice gas automata idea. We choose that the time step is small, and it is this parameter which we use for our asymptotic analysis. By choosing a discrete scheme such that the zero order dynamics give the Euler equations, we show in Section \ref{sec:discrete} that we retrieve the same computational system as the discrete time Boltzmann anyway. We pursue the dispersive dynamics as the higher order dynamics, in the time step, of the difference scheme that we have chosen. Such a perspective is motivated by, for example Goodman and Lax \cite{laxdispersive} where it was shown that a particular difference scheme applied to the partial differential equation,
\begin{equation}
\frac{\partial}{\partial t}u(x,t)  + u(x,t)\frac{\partial}{\partial x}u(x,t) = 0,
\end{equation}
recovers at the second order in the space difference parameter $\Delta$, the KdV equation,
\begin{equation}
\frac{\partial}{\partial t}u(x,t)  + u(x,t)\frac{\partial}{\partial x}u(x,t) + \frac{1}{6}\Delta^2 u(x,t)\frac{\partial^3}{\partial x^3}u(x,t) =0
\end{equation}
\par In tandem with the discrete time step asymptotics we hypothesize the existence of a slow Invariant Manifold\cite{invariantmanifolds}, we calculate the general form of this manifold and use it to find an expression for the macroscopic dynamics of general discrete time systems, with both discrete and continuous velocities. As we have stated our choice of the small parameter to be used in the asymptotics is the discrete time step itself $\epsilon$. The dynamics of the quasi-equilibrium approximation(the Maxwell distribution in the continuum) define the zero order dynamics, higher order dynamics are given by the correction to the equilibrium of the same order \cite{qestates}. We match the dynamics of the distribution function at microscopic and macroscopic \cite{shortmemory} levels to find an expression for the first order non-equilibrium component of the distribution function. Together the zero and first order components of the distribution are sufficient to calculate the macroscopic dynamics up to the first (dissipative) order. 
\par As well as deriving a general expression  of the macroscopic dynamics, we additionally provide two examples of both discrete and continuous velocity systems. We find that despite the qualitative difference between continuous and discrete time systems, in discrete time we can still recover the Navier-Stokes equations in a continuous velocity system. In the discrete velocity system, however, the discretizations examined display, as expected, additional errors in the dissipative part. The precise form of these errors is subject to the discrete velocity set chosen and given for the two common examples we use. 
\par Finally we test, by a short wave perturbation, the stability of the dynamics of the discrete velocity system. 

\section{Background and Notation}
We begin with a short summary of the background to this work which will introduce some of the requisite ideas and notation. This brief discussion will be sufficient for this work, for many more details regarding the background and theory to Boltzmann systems a large number of works exist by several authors \cite{succitlbe,karlince,luoce,mesoscale}. 
\par The Boltzmann Equation is concerned with the time evolution of what might be described physically as the density of
particles. This density function is denoted $f$ and is a function of space, velocity space and time $f \equiv f(x, v, t)$. It is given as
follows;
\begin{equation}
\frac{\partial}{\partial t} f + v \cdot \nabla_x f  = Q^c(f).
\label{eq:contboltzmann}
\end{equation}
Already we need a certain amount of details to formalize what we have written. We begin with the space variable $x$, this
is a smooth $k$-dimensional manifold, for example Euclidean space, or a torus, in finite dimensions. The velocity variable $v$
ranges over the same space. In the continuous time system above $t$ is a single real variable, however we will dispense with
this very soon. The final notation to mention in \ref{eq:contboltzmann} is the collision integral $Q^c$, a differentiable
transformation of the population function. More detailed properties which we require of this function will be given later.
\par In fact we are not at all concerned with the continuous time Boltzmann evolution, but the corresponding discrete time
system:
\begin{equation}
f(x + \epsilon v, v, t + \epsilon) = f(x,t) + Q(f(x,t)).
\label{eq:discreteboltzmann}
\end{equation}
This is the discrete time Boltzmann system which is at the core of this work. In contrast with the continuous time system
we have introduced the time step $\epsilon$, which will play a key role in our analysis. The time step should be small and it restricts
admissible values of time to the subgroup ${\epsilon Z \in R}$.
\par The left two terms in Eq. \ref{eq:discreteboltzmann} are collectively termed advection or free flight, if the collision integral is omitted, the exact
evolution of the population function at a point may be written as $f(x, v, t+\epsilon) = f(x- \epsilon v, v, t)$. The physical interpretation is
that particles move freely under their own momentum with no interaction between themselves. For our analysis we can use
the fact that advection is smooth and we will be able to take a Taylor series expansion to any finite order that we need. The
rightmost term is again called the collision integral and is denoted slightly differently to notate that it may differ somewhat
in form from its continuous counterpart.
\par Rewriting the right hand side of Eq. \ref{eq:discreteboltzmann} leads us to a perspective which will be key to our approach. Rewriting the collision
part as $F(f) = f + Q(f)$, allows us to present Eq. \ref{eq:discreteboltzmann} differently:
\begin{equation}
f(x, v, t + \epsilon) = F(f(x - \epsilon v, v, t)). 
\label{eq:superposition}
\end{equation}
In this way, the discrete time Boltzmann system can then be thought of as a superposition of the advection and collision operations.
Advection and collision steps happen in turn, this is qualitatively different from the continuous time system where both
operations might be said to be happening simultaneously at all times.
\par The evolution of the particle density function $f$ can be said to describe the microscopic evolution of the system. To
recover the macroscopic system we need to sum accross the velocity space. Here then is the key distinction between the
continuous $(v \in \mathbb{R}^k)$ and discrete $(v \in \{v^\alpha\} \alpha = 0 . . . n)$ velocity systems. In the discrete velocity case, instead of considering
the density function to be a function of velocity, we define different density functions $f^\alpha$, one per velocity vector and we can
denote then by $f$ the ordered set of them $f \equiv {f^\alpha(x, t), \; \alpha = 0,\ldots ,n}$. For example the density of the fluid is calculated in the
continuous and discrete schemes respectively as
\begin{equation}
\begin{array}{cc}
\rho(x, t) = \int_{\mathbb{R}^k} f(x, v, t) dv, & \rho(x, t) =\sum_{\alpha=0}^n f^\alpha(x, t)
\end{array}
\label{eq:contdiscmoments}
\end{equation}
In fact which macroscopic variables $M$ we use are not important to our initial analysis. The properties that are important
for the macroscopic moments are firstly that the operator $m$, which recovers the macroscopic moments from the density
function $f$ $(m(f) = M)$ is linear. The second property is that these moments are invariant under the collision operation,
that is $m(f) = m(F(f))$.
\par Defining the macroscopic variables allows us to discuss another property of the collision operation $F$. The quasi-equilibrium
is the unique vector $f = f^{\mathrm{eq}}_M$ such that $F(f^{\mathrm{eq}}_M) = f^{\mathrm{eq}}_M$ and $m(f^{\mathrm{eq}}_M) = M.$ Existence and uniqueness of this quasi-equilibrium will be assumed in this analysis. There is exactly one equilibrium point per value of M and together they form the quasi-equilbrium
manifold through the space of the density function.

\section{The discretization of Velocity Space}
\label{sec:discrete}
While we calculate the dynamics of the continuous velocity system later, they serve only for the purpose of calculating the
error incurred by an approximation. For practical computations we would use a discrete velocity system. What we have
control of is which discrete approximation to use. Two different approaches will lead us to the same result.
\par Firstly we consider an actual discretization of the Maxwell-Boltzmann distribution in $D$ dimensions, that is the known quasi-equilibrium in continuous velocity space.
\begin{equation}
f^{\mathrm{eq}} = \rho (2 \pi T)^{-\frac{D}{2}}\exp\left(\frac{-(v-u)^2}{2T}\right).
\end{equation}
The $2+D$ thermodynamic moments here are the density $\rho$, momentum $u$ and temperature $T$ . This distribution is multiplied
with low order polynomials of the velocity and integrated to retrieve the macroscopic moments.
\begin{equation}
\begin{split}
&\int_{\mathbb{R}^D}f^{\mathrm{eq}} dv = \rho , \\
&\int_{\mathbb{R}^D}vf^{\mathrm{eq}} dv = \rho u, \\
&\int_{\mathbb{R}^D}v^2f^{\mathrm{eq}} dv = \rho DT + \rho u^2.
\end{split}
\end{equation}
We can view the velocity set as a quadrature approximation to these integrals \cite{heluo}. With such a perspective the choice of which
quadrature to use may seem obvious, integrals of Gaussian type (as are the moment integrals) can be integrated exactly
using a Gauss-Hermite quadrature. Unfortunately several stumbling blocks prevent us from performing this exact integration.
\par The first problem is the necessary change of variable in the equilibrium distribution. For a Hermitian quadrature of any
order, the nodes should be distributed symmetrically about the centre of the Gaussian (in this case $u$), and the coordinate of integration should be normalized. Effectively it is necessary to apply a change of variable to the moment integrals so that the exponential term  is of the simple form $\exp(-v^2)$. Unfortunately, to do this in practice would require us to know both $u$ and $T$ before performing the integration. Of course we may still be
able to solve such a system, perhaps by some iterative method (moving the quadrature nodes) up to an arbitrary
degree of accuracy, but only if we can evaluate the density function anywhere we choose. In practice of course we can only
store a finite number of evaluations of the density function, and these points of evaluation are chosen pre-emptively to be the
same across all lattice sites and time steps (in order that advection is an exact operation). Because of these factors we should
not expect to be able to integrate exactly a density function of Gaussian type with a general momentum and temperature.
\par The popular method to partially rectify this problem is to assume that $u$ is close to zero. If we choose to work in an
athermal system (where $T$ is some constant) then we can exand the Maxwell distribution about the point $u = 0$ up to the
second order giving us
\begin{equation}
f^{\mathrm{eq}} = \rho (2 \Pi T)^\frac{-D}{2}\exp\left(\frac{-v^2}{2T}\right)\left(1+\frac{vu}{T} - \frac{(T-v^2)u^2}{2T^2}\right)
\label{eq:expandedmaxwell}
\end{equation}
The second order expansion is taken to get sufficient terms that the temperature moment integral is calculated correctly.
With a constant $T$ , and $u$ no longer affecting the midpoint of the distribution, the same nodes and weights can be used for
all space points and time steps. Assuming the proper transformation of variable these nodes and weights are the standard hermitian ones.
\par The alternative method of defining the quasi-equilibrium is by solution of a linear system. Since the moments are
calculated by linear combinations of the discrete quasiequilibria, these equilibria can be found by the inversion of a matrix of
components of the discrete velocities, to illustrate this we give an extremely simple example. We consider an athermal one
dimensional system with three discrete velocities $\{-1, 0, 1\}$ and fixed temperature $1/3$. We create a matrix where each line
represents an elementwise power of the velocities (up to the second order) and the corresponding vector of moments we desire.
Together this creates the following linear system.
\begin{equation}
\left[ \begin{array}{ccc}
1 & 1 & 1 \\
-1 & 0 & 1 \\
1 & 0 & 1 
\end{array}
\right] 
\cdot f^{\mathrm eq} = \left[\begin{array}{c} \rho \\ \rho u \\ \rho /3 + \rho u^2 \end{array}\right]
\end{equation}
Solution of the above system yields a vector of equilibria
\begin{equation}
f^{\mathrm{eq}} =\left\{ \frac{1}{6}\left(\rho - 3 \rho u + 3\rho u^2\right),\frac{2}{3}\left(\rho - \frac{3}{2}\rho u^2\right),\frac{1}{6}\left(\rho + 3 \rho u + 3\rho u^2\right) \right\}
\end{equation}
Inspection reveals this to be exactly the same, with weights pre-included, as the discretized Taylor approximation to the
Boltzmann (Eq. \ref{eq:expandedmaxwell}) where the same velocity set is applied. We note that in this example we have exactly the necessary
amount of velocities (that is one per moment) and hence have a square matrix. As in the case of any linear system, were
we to have fewer velocities it would become likely that the system would become unsolvable. For each additional discrete
velocity that we add above the necessary amount, we can impose an additional linear constraint. Often the choice would be
to zero higher order moments of the equilibrium distribution, that is enforcing that the sum of the equilibria multiplied with
polynomials of the discrete velocity components, with order higher than two, should be zero. Another option is to create additional, non-hydrodynamic moments in order to suppress instabilities \cite{nonhydro}.

\section{Invariant Manifolds for Discrete Time Boltzmann Systems}
We will, by finding a general form for an invariant manifold, calculate general microscopic dynamics for discrete time Boltzmann systems. To do this we need to make some assumptions regarding the stability of collisions and the smoothness of the distribution function.
\par For the next subsections we introduce the notation that fields of distribution functions of moments will be written in gothic font, hence the field of macroscopic variables, for example, will be denoted $\mathfrak{M}$.
\par We assume that collisions are stable, and for any admissible initial
	state $f$ iterations $F^p(f)$ converge to unique
	equilibrium point $f^{\rm eq}$ exponentially fast and
	uniformly:
	\begin{equation}\label{expstabuni}
	\|F^p(f) - f^{\rm eq}_{m(f)}\| < C \exp(-
	\lambda p)\|f - f^{\rm eq}_{m(f)}\|,
	\end{equation}
	where the Lyapunov exponent $\lambda > 0$ and pre-factor $C>0$ are
	the same for all admissible $f$.
In the limit $\epsilon = 0$ there is no free flight, the field of
macroscopic variables $\mathfrak{M}$ does not change, and the field
of distributions $\mathfrak{f}$ converges to the local equilibrium field $\mathfrak{f}^{\rm
eq}_M$ by repeated application of the collision operation. Since each collision occurs instantaneously, the superposed collisions become a projection onto the local equilibrium in zero time.

In order to discuss small $\epsilon > 0$ we need to evaluate the 
change of macroscopic variables in free flight during time step
$\epsilon$. To find the $q$th term of the non-equilibrium density function of a discrete velocity system, we make two
assumptions:
\begin{itemize}
\item{The first $q$ derivatives of $f_M$ along vector fields $v^{\alpha}$
exist and are bounded.}
\item{The differentials $(D_M^{(q)}f^{\rm eq}_M)$,  are
uniformly bounded for all $M$.}
\end{itemize}
\par Our expression for the manifold will be of the form of an aysmptotic expansion in the small parameter $\epsilon$,
\begin{equation}
f^{\text{inv}} = f^{(0)} + \epsilon f^{(1)} + \epsilon f^{(2)} + o(\epsilon).
\label{eq:IVManifold}
\end{equation} 
Our first goal is to find a prescription for this $f^{(1)} $ term. The zero order term of this is simply given by the quasi-equilibrium distributions, that is $f^{(0)}  \equiv f^{\text{eq}} $. For the first order  term we will take expansions of the distribution function in terms of time and of moments and equate them. That is for each order in epsilon we can take the effect of a complete LBM step (advection and collision) and match the effect on the distribution function to that of taking the Taylor approximation of the manifold through the moments up to the same order. In other words we match the dynamics of the microscopic and macroscopic scales on an order by order basis.

\subsection{The Invariance Equation}
\index{Invariance equation}
The procedure we use can also be described in terms of the invariant manifold hypothesis. Coupled steps of advection and collision form a chain of states of the population function belonging to the manifold. Since the number of discrete velocities used is normally larger than the number of macroscopic moments, there are an infinite number of possible population distributions which can give rise to the same configuration of moments, however only one of these distributions exists on the manifold. We use a Taylor approximation to the manifold and match it with a single coupled step to find the components, at different orders of the time step, of the distribution function.
\par If we consider $\mathfrak{f}_{\mathfrak{M}}^{\text{inv}}$ to be the field of population distributions on the manifold with corresponding field of macroscopic moments $\mathfrak{M}$ then in a continuous time system this invariance property can be defined as
\begin{equation}
D_t \mathfrak{f}_{\mathfrak{M}}^{\text{inv}}= D_\mathfrak{M} \mathfrak{f}_{\mathfrak{M}}^{\text{inv}} \cdot D_t \mathfrak{M}.
\end{equation}
Here the derivative $D_\mathfrak{M}$ indicates the derivative through the field of macroscopic moments $\mathfrak{M}$, which the field of distributions on the manifold $\mathfrak{f}_{\mathfrak{M}}^{\text{inv}}$ is parameterized by. Altogether the rate of change of the population function is equal to the rate of change of the moments multiplied by the change of the populations with respect to the moments. The discrete time analogy of this is given by,
\begin{equation}
\left(\mathfrak{f}^{\text{inv}}_{\mathfrak{M}}\right)' = \mathfrak{f}^{\text{inv}}_{\mathfrak{M}'}.
\label{eq:invariance}
\end{equation}
where the prime notates the next time step, therefore the left hand side of this equation is given by Eq. \ref{eq:superposition}.
\par  From now on we dispense with the gothic notation for fields of variables and simply use a standard font to describe distribution functions or macroscopic variables.

\subsection{The expansion of the distribution function following a step in the LBM chain}
A little abuse of notation will make the same calculations in this section both brief and meaningful. No distinction will be
made here between continuous and discrete velocity systems, however implicitly there is one. As previously $f \equiv f(x, v, t)$
for the continuous case, and $f \equiv {f^\alpha(x, t), \; \alpha = 0,\ldots ,n}$ in the discrete case. For the purposes of our analysis, in both the continuous and discrete velocity systems the space coordinate $x$ is continuous. Due to this, even though we may choose to only evaluate $f$ at a discrete number of lattice sites, the space derivative $D_x f$ remains well defined, even for the discrete velocity case. The notation is similar for moment derivatives, where $D_M f_M$ refers to the change in moments of the function $f$ evaluated at the moments given by the subscript to $f$. When the moment operator $m$ is applied this refers to the integral in the continuous case and the summation in the discrete case as given in Eq. \ref{eq:contdiscmoments}.
\par With these ideas in mind the procedure below is valid for both cases. The first ingredient for the time step expansion is the Taylor series of the advection operation up to the required order in $\epsilon$. For the first order we have
\begin{equation}
f(x-\epsilon v) = f - \epsilon v \cdot D_x f + o(\epsilon).
\end{equation}
Combining this with (\ref{eq:IVManifold}) we have to the first order,
\begin{equation}
f(x-\epsilon v) = f^{(0)} - \epsilon v \cdot D_x f^{(0)} + \epsilon f^{(1)} + o(\epsilon).
\end{equation}
Applying a collision operation gives the complete, composite discrete time step,
\begin{equation}
\left(f^{\text{inv}}_{M}\right)' = F\left(f^{(0)} - \epsilon v \cdot D_x f^{(0)} + \epsilon f^{(1)} + o(\epsilon)\right).
\label{eq:toexplainlinear}
\end{equation}
The second ingredient is to use a linearised version of the collision operation, this is sufficient to get the first order populations correctly. Here the linearisation is made about the equilibrium corresponding to the populations to be collided,
\begin{equation}
f  \mapsto f_{m(f)}^{\text{eq}} + \left(D_fF\right)_{f_{m(f)}^{\text{eq}}}\left(f - f_{m(f)}^{\text{eq}} \right).
\label{eq:linearizedcollision}
\end{equation}
Due to the linearity we can move the error term in Eq \ref{eq:toexplainlinear} outside the collision altogether. The linearisation is then made about the equilibrium defined by the moments of the first order advected populations
\begin{equation}
M_1' = m\left(f_{M}^{(0)} - \epsilon v\cdot D_xf_M^{(0)} \right) = M + m\left( - \epsilon v \cdot D_xf_{M}^{(0)} \right),
\label{eq:momentsfirstorder}
\end{equation}
Finally then for the first order approximation to the next step through the time step expansion we have,
\begin{equation}
\begin{split}
\left(f^{\text{inv}}_{M}\right)'  = f_{M_1'}^{\text{eq}}
 + \left(D_fF\right)_{f_{M_1'}^{\text{eq}}} \left(f_{M}^{(0)} + \epsilon f_{M}^{(1)}  - \epsilon v \cdot D_xf_{M}^{(0)}   -  f_{M_1'}^{\text{eq}} \right) +  o(\epsilon).
\end{split}                      
\label{eq:finallhs}
\end{equation}
\subsection{The expansion of the invariance equation following a time step}
With the expansion of the left hand of (\ref{eq:invariance}) complete we consider the right hand side. Here we find the Taylor expansion of the invariant manifold up to the linear term so,
\begin{equation} 
f_{M'} =f_{M} + (D_M f_{M}) \cdot m\left( - \epsilon v \cdot D_xf_{M}  \right)  + o(\epsilon).
\label{eq:rhsdm}
\end{equation}
Substituting (\ref{eq:IVManifold}) into (\ref{eq:rhsdm}) we have
\begin{equation} 
f_{M'} = f_{M}^{(0)} + \epsilon f_{M}^{(1)} + (D_M f_{M}^{(0)})\cdot m\left( - \epsilon v \cdot D_xf_{M}^{(0)}\right)  + o(\epsilon).
\label{eq:finalrhs}
\end{equation}
We can now equate (\ref{eq:finallhs}) and (\ref{eq:finalrhs}) for a first order approximation to (\ref{eq:invariance}),
\begin{multline}
f_{M_1'}^{\text{eq}}
 + \left(D_fF\right)_{f_{M_1'}^{\text{eq}}}\left(f_{M}^{(0)} + \epsilon f_{M}^{(1)}  - \epsilon v \cdot D_xf_{M}^{(0)}   -  f_{M_1'}^{\text{eq}} \right) \\ = f_{M}^{(0)} + \epsilon f_{M}^{(1)} + (D_M f_{M}^{(0)})\cdot m\left( - \epsilon v \cdot D_xf_{M}^{(0)}\right).
\label{eq:f1equation1}
\end{multline}
Of course in a similar style to (\ref{eq:rhsdm}),
\begin{equation}
f_{M_1'}^{\text{eq}} = f_{M}^{(0)} + (D_M f_M^{(0)}) \cdot m\left( - \epsilon v \cdot D_xf_{M}^{(0)}  \right).
\end{equation}
Substituting back into (\ref{eq:f1equation1}) we have 
\begin{equation}
\left(D_fF\right)_{f_{M_1'}^{\text{eq}}}\left(\epsilon f_{M}^{(1)} - \epsilon v \cdot D_xf_{M}^{(0)} - (D_M f_{M}^{(0)})\cdot m\left( - \epsilon v \cdot D_xf_{M}^{(0)}  \right) \right)  = \epsilon f_{M}^{(1)} 
\label{eq:f1equation2}
\end{equation}
This equation forms the prototype to find $f_{M}^{(1)}$ for different possible collision operations, it implicitly gives the first order approximation to the invariance equation (\ref{eq:invariance}). It depends on the choice of the velocity set, the quasiequilibrium and the collision integral.
\subsection{Example First Order Invariant Manifolds}
We consider two possible examples of collisions. The first example is the simple Ehrenfest step \cite{ehrenfest},
\begin{equation}
F\left( f \right) = f_{m(f)}^{\text{eq}}
\end{equation}
we immediately have,
\begin{equation}
\left(D_fF\right)_{f_{m(f)}}^{\text{eq}}\left( f \right) = 0.
\end{equation}
Substituting back into (\ref{eq:f1equation2}),
\begin{equation}
f_{M}^{(1)} = 0.
\end{equation}
This of course expected since using Ehrenfests steps for the collisions we should expect to return at every time step to the quasi-equilibrium manifold.
\par
\begin{figure}[h!]
\centering
\includegraphics[width=1\textwidth]{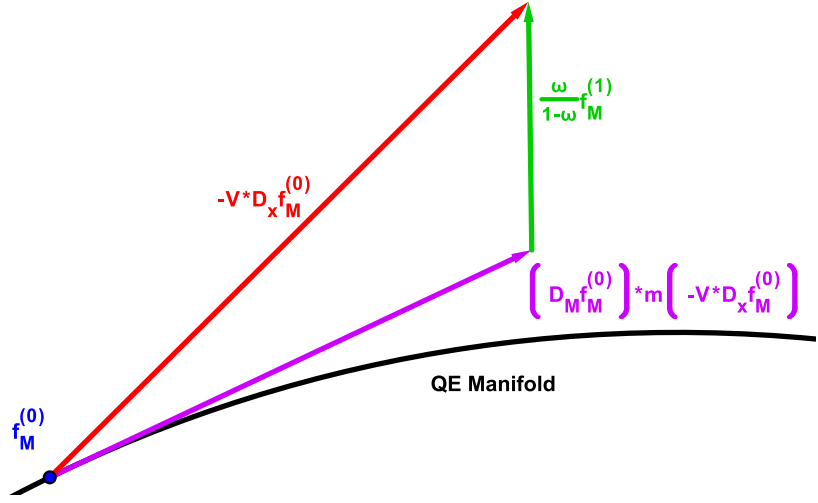}
\caption{Graphical representation of the $f_{M}^{(1)}$ for the BGK collision. Adding the $f_{M}^{(1)}$ term to the quasiequilibrium manifold gives the invariant manifold to first order in $\epsilon$. In particular the collision parameter $\omega$ is critical, for $\omega > 1$ the direction of the $f_{M}^{(1)}$ term is inverted and consequently the invariant manifold is below (in the sense of this illustration) the quasiequilibrium. Therefore at each step the advection operation crosses the quasiequilibrium and the collision returns below it.}
\label{fig:bgkf1}
\end{figure}

 The second example of a collision operator is the \index{BGK collision}BGK collision \cite{BGK},
\begin{equation}
F\left( f \right) = f + \omega\left( f_{m(f)}^{\text{eq}} -  f \right).
\end{equation}
Differentiating we have
\begin{equation}
\left(D_fF\right)_{f_{m(f)}^{\text{eq}}}\left( f \right) = \left( 1 - \omega\right)\cdot f
\end{equation}
Substituting this into (\ref{eq:f1equation2}),
\begin{equation}
\left( 1 - \omega\right)\cdot\left(\epsilon f_{M}^{(1)} - \epsilon v \cdot D_xf_{M}^{(0)}   - (D_M f_{M}^{(0)}) \cdot m\left( - \epsilon v \cdot D_xf_{M}^{(0)}  \right) \right) = \epsilon f_{M}^{(1)}. 
\label{eq:f1bgk1}
\end{equation}
We can multiply out (\ref{eq:f1bgk1}) and solve for, $f_{M}^{(1)}$.
\begin{equation}
\frac{\omega}{1 - \omega}f_{M}^{(1)} = \left(  -v \cdot D_xf_{M}^{(0)} \right) - (D_M f_{M}^{(0)}) \cdot m\left( - v \cdot D_xf_{M}^{(0)}  \right) 
\label{eq:f1bgk}
\end{equation}
Figure \ref{fig:bgkf1} graphically demonstrates the $f_{M}^{(1)}$ for the BGK collision type. In particular for this example there is a critical parameter value at $\omega = 1$. For $\omega = 1$ we recover the Ehrenfest step, for $\omega > 1$ we have the normal BGK over-relaxation where both of the coupled steps of advection and collision cross the quasiequilibrium manifold, one in each direction.

\subsection{Second Order Manifolds and an Example}
The next goal is to find an equation analagous to (\ref{eq:f1equation2}) for the second order term of the invariant manifold. During the next section we use a linear collision operation, in this case the linearised collision we use produces the exact same result as the original collision. We restart the procedure using second order expansions where appropriate, the first of these is the Taylor expansion of the advected populations.
\begin{equation}
f(x - \epsilon v) = f - \epsilon v\cdot D_xf 
+ \frac{\epsilon^2}{2}v\cdot D_x\left(v\cdot D_xf\right) + o(\epsilon^2)
\label{eq:taylorexpansion2}
\end{equation}
The second order population expansion is also used.
\begin{equation}
f^{\text{inv}}  = f^{(0)} + \epsilon f^{(1)} + \epsilon^2 f^{(2)} + o(\epsilon^2).
\label{eq:populationexpansion2}
\end{equation}
Altogether the second order expansion of the advected populations is,
\begin{multline}
f(x - \epsilon v) = f^{(0)} - \epsilon v\cdot D_xf^{(0)} 
+ \frac{\epsilon^2}{2}v\cdot D_x\left(v\cdot D_xf^{(0)}\right) \\ + \epsilon f^{(1)}  - \epsilon^2v\cdot D_xf^{(1)} +\epsilon^2f^{(2)} + o(\epsilon^2)
\label{eq:taylorppop2}
\end{multline}
We use the linearized collision integral in replacement of the original collision operation,
\begin{multline}
(f_M^{\text{inv}})' = f_{M_2'}^{\text{eq}}
 + \left(D_fF\right)_{f_{M_2'}^{\text{eq}}}\Big(f_M^{(0)} - \epsilon v\cdot D_xf_M^{(0)}   +  \frac{\epsilon^2}{2}v\cdot D_x\left(v\cdot D_xf_M^{(0)}\right)  \\ + \epsilon f_M^{(1)} - \epsilon^2v \cdot D_xf_M^{(1)} +\epsilon^2f_M^{(2)}   -  f_{M_2'}^{\text{eq}} \Big) +  o(\epsilon^2).
\label{eq:lincollptaylor}
\end{multline}
where $M_2'$ are the moments of the post advection populations to \emph{second} order,
\begin{equation}
M_2' = m\left(f_M^{(0)} - \epsilon v \cdot D_xf_M^{(0)} +  \frac{\epsilon^2}{2}v\cdot D_x\left(v\cdot D_xf_M^{(0)}\right)  - \epsilon^2v \cdot D_xf_M^{(1)} \right),
\end{equation}
For the right hand side of (\ref{eq:invariance}) we use a second order approximation to the invariant manifold,
\begin{equation} 
f^{\text{inv}}_{M'} = f_M^{\text{inv}}(\mathbf{x},v_\alpha) + \Delta M_2 \cdot D_M f_M^{\text{inv}} 
+ \frac{1}{2}\Delta M_2 \cdot D_M (\Delta M_2 \cdot D_M f_M^{\text{inv}}) + o((\Delta M_2)^2).
\label{eq:rhsdm2}
\end{equation}
where for berevity $\Delta M_2$ is the difference between the moments of the post and pre advection populations to second order,
\begin{equation}
\Delta M_2 = m\left( - \epsilon v \cdot D_xf_M^{(0)} + \frac{\epsilon^2}{2}v\cdot D_x\left(v\cdot D_xf_M^{(0)}\right)- \epsilon^2v \cdot D_xf_M^{(1)} \right) = M_2' - M.
\label{eq:DeltaM2}
\end{equation}
Substituting (\ref{eq:populationexpansion2}) and (\ref{eq:DeltaM2}) into (\ref{eq:rhsdm2}) we have
\begin{equation}
\begin{split} 
& f^{\text{inv}}_{M'}=  f_M^{(0)} + \epsilon f_M^{(1)} + \epsilon^2 f_M^{(2)} + m\Big( - \epsilon v \cdot D_xf_M^{(0)} + \frac{\epsilon^2}{2}v\cdot D_x\left(v\cdot D_xf_M^{(0)}\right) 
\\ & \qquad \qquad - \epsilon^2v \cdot D_xf_M^{(1)} \Big)\cdot D_M f_M^{(0)} \\ & \qquad + \frac{1}{2}m\left( - \epsilon v\cdot D_xf_M^{(0)}\right)\cdot D_M\left( m\left( - \epsilon v\cdot D_xf_M^{(0)}\right)\cdot D_M f_M^{(0)}\right) \\ & \qquad + \epsilon m\left( - \epsilon v \cdot D_xf_M^{(0)}\right)\cdot D_M f_M^{(1)} + o(\epsilon^2).
\label{eq:finalrhs2}
\end{split}
\end{equation}
With the expansions of both sides complete we can equate(\ref{eq:lincollptaylor}) and (\ref{eq:finalrhs2}),
\begin{equation}
\begin{split}
& f_{M_2'}^{\text{eq}} + \left(D_fF\right)_{f_{M_2'}^{\text{eq}}}\Big(f_M^{(0)} - \epsilon v\cdot D_xf_M^{(0)}  + \frac{\epsilon^2}{2}v\cdot D_x\left(v\cdot D_xf_M^{(0)}\right) + \epsilon f_M^{(1)} \\ & \qquad - \epsilon^2v \cdot D_xf_M^{(1)} +\epsilon^2f_M^{(2)}   -  f_{M_2'}^{\text{eq}} \Big) \\ & = f_M^{(0)} + \epsilon f_M^{(1)} + \epsilon^2 f_M^{(2)}  \\ & \qquad + m\left( - \epsilon v \cdot D_xf_M^{(0)} + \frac{\epsilon^2}{2}v\cdot D_x\left(v\cdot D_xf_M^{(0)}\right) - \epsilon^2v \cdot D_xf_M^{(1)} \right)\cdot D_M f_M^{(0)} \\ & \qquad +\frac{1}{2} m\left( - \epsilon v\cdot D_xf_M^{(0)}\right)\cdot D_M\left( m\left( - \epsilon v\cdot D_xf_M^{(0)}\right)\cdot D_M f_M^{(0)}\right)  \\ & \qquad + \epsilon m\left( - \epsilon v \cdot D_xf_M^{(0)}\right)\cdot D_M f_M^{(1)}
\label{eq:f2equation1}
\end{split}
\end{equation}
Analagously to the first order case we note that,
\begin{equation}
\begin{split} 
f^{\text{eq}}_{M_2'}= & f_M^{(0)} + m\Big( - \epsilon v \cdot D_xf_M^{(0)} +\frac{\epsilon^2}{2}v\cdot D_x\left(v\cdot D_xf_M^{(0)}\right)  - \\ & \qquad   \epsilon^2v \cdot D_xf_M^{(1)} \Big)\cdot D_M f_M^{(0)} \\ &+ m\left( - \epsilon v\cdot D_xf_M^{(0)}\right)\cdot D_M\left( m\left( - \epsilon v\cdot D_xf_M^{(0)}\right)\cdot D_M f_M^{(0)}\right).
\end{split}
\end{equation}
Substituting this back into (\ref{eq:f2equation1}) we have the final prototype for $f_M^{(2)}$ which this time is given implicitly,
\begin{multline}
\left(D_fF\right)_{f_{M_2'}^{\text{eq}}}\Bigg( - \epsilon v\cdot D_xf_M^{(0)}  + \frac{\epsilon^2}{2}v\cdot D_x\left(v\cdot D_xf_M^{(0)}\right) + \epsilon f_M^{(1)} - \epsilon^2v \cdot D_xf_M^{(1)}  +\epsilon^2f_M^{(2)}\\ - m\left( - \epsilon v \cdot D_xf_M^{(0)} +\frac{\epsilon^2}{2}v\cdot D_x\left(v\cdot D_xf_M^{(0)}\right)  - \epsilon^2v \cdot D_xf_M^{(1)} \right)\cdot D_M f_M^{(0)} \\- \frac{1}{2}m\left( - \epsilon v\cdot D_xf_M^{(0)}\right)\cdot D_M\left( m\left( - \epsilon v\cdot D_xf_M^{(0)}\right)\cdot D_M f_M^{(0)}\right)  \Bigg) \\  =  \epsilon f_M^{(1)} + \epsilon^2 f_M^{(2)} + \epsilon m\left( - \epsilon v \cdot D_xf_M^{(0)}\right)\cdot D_M f_M^{(1)}
\label{eq:f2equation2}
\end{multline}
We return to the BGK collision for a specific example of an $f_M^{(2)}$ term. (\ref{eq:f2equation2}) becomes,
\begin{multline}
(1-\omega)\cdot\Bigg( - \epsilon v\cdot D_xf_M^{(0)}  + \frac{\epsilon^2}{2}v\cdot D_x\left(v\cdot D_xf_M^{(0)}\right) + \epsilon f_M^{(1)} - \epsilon^2v \cdot D_xf_M^{(1)}  +\epsilon^2f_M^{(2)}\\ - m\left( - \epsilon v \cdot D_xf_M^{(0)} +\frac{\epsilon^2}{2}v\cdot D_x\left(v\cdot D_xf_M^{(0)}\right)  - \epsilon^2v \cdot D_xf_M^{(1)} \right)\cdot D_M f_M^{(0)} \\- \frac{1}{2}m\left( - \epsilon v\cdot D_xf_M^{(0)}\right)\cdot D_M\left( m\left( - \epsilon v\cdot D_xf_M^{(0)}\right)\cdot D_M f_M^{(0)}\right)  \Bigg) \\  =  \epsilon f_M^{(1)} + \epsilon^2 f_M^{(2)} + \epsilon m\left( - \epsilon v \cdot D_xf_M^{(0)}\right)\cdot D_M f_M^{(1)}
\end{multline}
Rearranging  and equating terms with $\epsilon$ order 2 gives us,
\begin{equation}
\begin{split}
\frac{\omega}{1 - \omega} f_M^{(2)}  = &  \frac{1}{2}v\cdot D_x\left(v\cdot D_xf_M^{(0)}\right) - v \cdot D_xf_M^{(1)}  \\ & - m\left( \frac{1}{2}v\cdot D_x\left(v\cdot D_xf_M^{(0)}\right)  - v \cdot D_xf_M^{(1)} \right)\cdot D_M f_M^{(0)} \\ & - \frac{1}{2}m\left( - v\cdot D_xf_M^{(0)}\right)\cdot D_M\left( m\left( - v\cdot D_xf_M^{(0)}\right)\cdot D_M f_M^{(0)}\right)\\ & - 
\frac{1}{1 -\omega}m\left( - v \cdot D_xf_M^{(0)}\right)\cdot D_M f_M^{(1)}
\end{split}
\end{equation}
We will not use these populations in the examples of macroscopic dynamics which we calculate in the next section. For the dissipative dynamics the first order populations are sufficient. We expect that this second order part should give rise to macroscopic dynamics equating to the Burnett equations in a continuous velocity system, with some additional error terms if a discrete velocity set is used.

\section{Macroscopic Equations}
In this section we are concerned with deriving equations for the macroscopic dynamics arising from several different example lattices. We expect that the lattice parameter $\epsilon$ should partly govern these dynamics and that the 1st order macroscopic dynamics should be governed by the 1st order population functions.
\par In order to find these dynamics we project the microscopic flow (advection) up to the required order, following one time step, onto the invariant manifold up to the same order \cite{invariantmanifolds,coarsegraining}. 
\par We can immediately perform a Taylor expansion in time on the macroscopic dynamics,
\begin{equation}
M' = M + \epsilon \frac{\partial}{\partial t}M+ o(\epsilon)
\end{equation}
We expect that the final model should be given in terms of a time derivative of the moments, we write this in a power series in terms of $\epsilon$,
\begin{equation}
\frac{\partial}{\partial t}M = \Psi^{(0)} + \epsilon\Psi^{(1)} + o(\epsilon)
\end{equation}
Combining these two we have
\begin{equation}
M' = M + \epsilon\Psi^{(0)} + o(\epsilon)
\label{eq:macro1storder}
\end{equation}
Equating (\ref{eq:momentsfirstorder}) and (\ref{eq:macro1storder}) we have
\begin{equation}
\Psi^{(0)}  = m\left( -v \cdot D_xf_{M}^{(0)} \right)
\label{eq:psi0eq}
\end{equation}
The corresponding second order approximation of the moments in time is
\begin{equation}
M' = M + \epsilon(\Psi^{(0)} + \epsilon \Psi^{(1)}) + \frac{\epsilon^2}{2}\frac{\partial}{\partial t}\Psi^{(0)} + o(\epsilon^2)
\label{eq:macro2ndorder}
\end{equation}
Equating terms on the second order of $\epsilon$ we have,
\begin{equation}
m\left(\frac{1}{2}v\cdot D_x\left(v\cdot D_xf_M^{(0)}\right)\right)+ m\left(- v \cdot D_xf_M^{(1)} \right)    = \Psi^{(1)}(t) + \frac{1}{2}\frac{\partial}{\partial t}\Psi^{(0)}(t)
\end{equation}
or
\begin{equation}
\Psi^{(1)}(t) = m\left(\frac{1}{2}v\cdot D_x\left(v\cdot D_xf_M^{(0)}\right)\right)+ m\left(- v \cdot D_xf_M^{(1)} \right) - \frac{1}{2}\frac{\partial}{\partial t}\Psi^{(0)}(t)
\label{eq:psi1eq}
\end{equation}
Although in this notation the evolution of the macroscopic moments of the continuous and discrete velocity systems can be
written in the same way, in practical examples we have no reason to suspect that they should be necessarily equivalent. In
the next two sections we calculate some example, equivalent, discrete and continuous systems, in order to compare them.
\section{Discrete Velocity Examples}
We will now demonstrate the exact first order dynamics of a popular choice of lattice scheme in one and two dimensions. The athermal schemes we consider are typically described in shorthand by the dimension within which they operate and the number of velocities used to form the lattice in the form DnQm, where $m$ and $n$ are integers representing the number of dimensions and velocities respectively. The general quasi-equilibrium  for these systems, including the two examples we use can be written in a general form
\begin{equation}
f^{\alpha,(0)}_M = W_\alpha \rho \left(1+ \frac{v^\alpha\cdot u}{c_s^2} + \frac{(v^\alpha\cdot u)^2}{2c_s^4} - \frac{u^2}{2c_s^2}\right)
\label{eq:polyquasiequil}
\end{equation}
This equilibrium defines an athermal system where the temperature is fixed. To complete the definition of the discrete system
requires only the selection of a velocity set and some accompanying weights $W_\alpha$.

\subsection{An athermal three velocity lattice (D1Q3)}
\label{sec:athermal3vel1d}
\par Our  1-D example lattice is one of the most common, the athermal 1-D lattice with 3 velocities. In this example the velocity vectors are $\left\{-1,0,1 \right\}$ and the speed of sound $c_s = 1/\sqrt{3}$ hence the equilbrium populations are derived from the general formula for athermal quasi-equilibria in any dimension where the additional parameters the weights $W_\alpha$ are $\left\{\frac{1}{6},\frac{2}{3},\frac{1}{6}\right\}$. 
\par For this case the populations are, 
\begin{equation}
\frac{1}{6}\left\{\rho(1-3u+3u^2,2\rho\left(1-\frac{3u^2}{2}\right),\rho(1+3u+3u^2) \right\}.
\end{equation}
For this lattice with unit distances we note that $v^4 = v^2$, $v^1 = v^3$ etc. We calculate the two components of $\Psi^{(0)}$ using the formulas for the moments. We have for the density derivative,
\begin{equation}
\Psi^{(0)}_1 = -\sum_\alpha v^\alpha  \frac{\partial}{\partial x}f_{M}^{\alpha,(0)}  
= - \frac{\partial}{\partial x}\sum_\alpha W_\alpha v^\alpha f_{M}^{\alpha,(0)} 
= -\frac{\partial}{\partial x} \rho u,
\end{equation}
and for the momentum derivative
\begin{multline}
\Psi^{(0)}_2 = - \sum_\alpha W_\alpha (v^\alpha)^2  \frac{\partial}{\partial x}f_{M}^{\alpha,(0)} 
= - \frac{\partial}{\partial x}\sum_\alpha W_\alpha (v^\alpha)^2 f_{M}^{\alpha,(0)} 
 \\ = -\frac{\partial}{\partial x}\left(\frac{\rho}{3} + \rho u^2\right).
\end{multline}
Now we examine the individual moments of the first order part in the case of the one dimensional lattice, as before we begin with the density,
\begin{equation}
\Psi^{(1)}_1 =  \frac{1}{2} \sum_{\alpha} (v^\alpha)^2 \frac{\partial^2}{\partial x^2}f_{M}^{\alpha,(0)}  - \sum_{\alpha} v^\alpha \frac{\partial}{\partial x}f_{M}^{\alpha,(1)} - \frac{1}{2}\frac{\partial}{\partial t}\Psi^{(0)}_1.
\end{equation}
The second term here is the space derivative of the momentum of the $f_{M}^{(1)}$ which equals zero due to all moments of non equilibrium components being zero and the first term can be calculated immediately from the quasi-equilibrium. Therefore,
\begin{equation}
\Psi^{(1)}_1 =   \frac{1}{2}\frac{\partial^2}{\partial x^2} \left(\frac{\rho}{3} + \rho u^2\right)  - \frac{1}{2}\frac{\partial}{\partial t}\Psi^{(0)}_1.
\end{equation}
The time derivative of $\Psi^{(0)}_1$ can be calculated by the chain rule,
\begin{equation}
\frac{\partial}{\partial t}\Psi^{(0)}_1 = \frac{\partial \Psi^{(0)}_1}{\partial \rho}\frac{\partial \rho}{\partial t} + \frac{\partial \Psi^{(0)}_1}{\partial \rho u}\frac{\partial \rho u}{\partial t}  = -\frac{\partial}{\partial x} \Psi^{(0)}_2 =  \frac{\partial^2}{\partial x^2} \left(\frac{\rho}{3} + \rho u^2\right)
\end{equation}
Substituting this back in we have,
\begin{equation}
\Psi^{(1)}_1 =  \frac{1}{2}\frac{\partial^2}{\partial x^2} \left(\frac{\rho}{3} + \frac{(\rho u)^2}{\rho}\right) - \frac{1}{2}\frac{\partial^2}{\partial x^2} \left(\frac{\rho}{3} + \rho u^2\right) = 0 .
\end{equation}
For the momentum moment we have
\begin{equation}
\Psi^{(1)}_2 =  \frac{1}{2} \sum_{\alpha} (v^\alpha)^3 \frac{\partial^2}{\partial x^2}f_{M}^{\alpha,(0)} - \sum_{\alpha} (v^\alpha)^2 \frac{\partial}{\partial x}f_{M}^{\alpha,(1)} - \frac{1}{2}\frac{\partial}{\partial t}\Psi^{(0)}_2.
\end{equation}
Recalling that that $v^3 = v^1$ we can simplify the first term so,
\begin{equation}
\frac{1}{2} \sum_{\alpha} (v^\alpha)^3 \frac{\partial^2}{\partial x^2}f_{M}^{\alpha,(0)} = \frac{1}{2} \sum_{\alpha} v^\alpha \frac{\partial^2}{\partial x^2}f_{M}^{\alpha,(0)} = \frac{1}{2} \frac{\partial^2}{\partial x^2} \rho u .
\end{equation}
For the second term we need to calculate the $f_{M}^{\alpha,(1)}$ terms. To do this we need to specify a collision type, we use the BGK collision described above (\ref{eq:f1bgk}).
\begin{multline}
\frac{\omega}{1-\omega} f_M^{(1)} = \frac{\partial}{\partial x}\left(2u\frac{\partial}{\partial x}\rho + \left(-2-3u^2\right)\frac{\partial}{\partial x}\rho u + 6u\frac{\partial}{\partial x}\rho u^2\right.,\\ -u\frac{\partial}{\partial x}\rho + \left(1+\frac{3}{2}u^2\right)\frac{\partial}{\partial x}\rho u -3u\frac{\partial}{\partial x}\rho u^2, \\ \left. 2u\frac{\partial}{\partial x}\rho + (-2-3u^2)\frac{\partial}{\partial x}\rho u + 6u\frac{\partial}{\partial x}\rho u^2 \right).
\end{multline}
This gives then
\begin{equation}
\sum_{\alpha} (v^\alpha)^2 \frac{\partial}{\partial x}f_{M}^{\alpha,(1)} =\frac{1-\omega}{\omega} \frac{\partial}{\partial x} \left(\frac{2}{3}u\frac{\partial}{\partial x}\rho + \left(-\frac{2}{3}-u^2\right)\frac{\partial}{\partial x}\rho u +2u\frac{\partial}{\partial x}\rho u^2 \right).
\end{equation}
Again using the chain rule,
\begin{multline}
\frac{\partial}{\partial t}\Psi^{(0)}_2 = \frac{\partial \Psi^{(0)}_2}{\partial \rho}\frac{\partial \rho}{\partial t} + \frac{\partial \Psi^{(0)}_2}{\partial \rho u}\frac{\partial \rho u}{\partial t}  = -\frac{\partial}{\partial x}\left(\frac{1}{3} - u^2\right)\Psi^{(0)}_1 - \frac{\partial}{\partial x} 2u \Psi^{(0)}_2 \\ = \frac{\partial}{\partial x}\left(\frac{2}{3}u\frac{\partial}{\partial x}\rho +\left(\frac{1}{3}- u^2\right)\frac{\partial}{\partial x}\rho u  + 2u\frac{\partial}{\partial x}\rho u^2\right).
\end{multline}
Substituting this all back in we have,
\begin{multline}
\Psi^{(1)}_2 =  \left(\frac{\omega-1}{\omega} -\frac{1}{2}\right)\frac{\partial}{\partial x}\left(\frac{2}{3}u\frac{\partial}{\partial x}\rho  +\left(-\frac{2}{3} -u^2\right)\frac{\partial}{\partial x}\rho u	+2u\frac{\partial}{\partial x}\rho u^2 \right)\\ = \frac{\omega-2}{2\omega}\frac{\partial}{\partial x}\left(\frac{2}{3}u\frac{\partial}{\partial x}\rho+\left( -\frac{2}{3}-u^2\right)\frac{\partial}{\partial x}\rho u	+2u\frac{\partial}{\partial x}\rho u^2 \right) \\ = \frac{\omega-2}{2\omega}\frac{\partial}{\partial x}\left(u^3\frac{\partial}{\partial x}\rho + \rho\left(3u^2-\frac{2}{3}\right)\frac{\partial}{\partial x}u \right) .
\end{multline}
The moment gradients are then to first order in $\epsilon$,
\begin{equation}
\begin{split}
&\frac{\partial}{\partial t}\rho = -\frac{\partial}{\partial x} \rho u \\
&\frac{\partial}{\partial t}\rho u = -\frac{\partial}{\partial x}\left(\frac{1}{3}\rho + \rho u^2\right)-\epsilon\frac{2-\omega}{2\omega}\frac{\partial}{\partial x}\left(u^3\frac{\partial}{\partial x}\rho + \rho\left(3u^2-\frac{2}{3}\right)\frac{\partial}{\partial x}u)\right)
\end{split}
\end{equation}

\subsection{An athermal nine velocity model (D2Q9)}
The 2-D example we consider is a popular 2d lattice consisting of 9 different velocities. If we identify $v_1$ as the horizontal component of a vector and $v_2$ the vertical component then the set of velocities is
\begin{multline}
v_\alpha = \left\{
\left( \begin{array}{c} 0 \\0 \end{array} \right),
\left( \begin{array}{c} 1 \\0 \end{array} \right),
\left( \begin{array}{c} 0 \\1 \end{array} \right),
\left( \begin{array}{c} -1 \\0 \end{array} \right),
\left( \begin{array}{c} 0 \\-1 \end{array} \right),
\right. \\ \left.
\left( \begin{array}{c} 1 \\1 \end{array} \right),
\left( \begin{array}{c} -1 \\1 \end{array} \right),
\left( \begin{array}{c} -1 \\-1 \end{array} \right),
\left( \begin{array}{c} 1 \\-1 \end{array} \right)
\right\}
\end{multline}
The equilibrium is then given by the polynomial formula (\ref{eq:polyquasiequil}) with corresponding weights
\begin{equation}
W_\alpha = \left\{ \frac{4}{9},\frac{1}{9},\frac{1}{9},\frac{1}{9},\frac{1}{9},\frac{1}{36},\frac{1}{36},\frac{1}{36},\frac{1}{36}\right\}
\end{equation}
\par As before we calculate the components of $\Psi^{(0)}$ using the formulas for the moments although this time we have two momentum density momentums for the two dimensions. We have for the density derivative,
\begin{equation}
\begin{split}
\Psi^{(0)}_1& = \sum_\alpha  \left( -v^\alpha \cdot  D_xf_{M}^{\alpha,(0)}\right) 
\\ & = - \frac{\partial}{\partial x_1}\sum_\alpha  v^\alpha_1 f_{M}^{\alpha,(0)} - \frac{\partial}{\partial x_2}\sum_\alpha  v^\alpha_2 f_{M}^{\alpha,(0)}
\\ & = -\frac{\partial}{\partial x_1} \rho u_1 - \frac{\partial}{\partial x_2} \rho u_2,
 \end{split}
\end{equation}
for the first momentum derivative
\begin{equation}
\begin{split}
\Psi^{(0)}_2& = \sum_\alpha  v^\alpha_1 \left( -v^\alpha \cdot D_xf_{M}^{\alpha,(0)}\right) 
\\ & = - \frac{\partial}{\partial x_1}\sum_\alpha  (v^\alpha_1)^2 f_{M}^{\alpha,(0)} - \frac{\partial}{\partial x_2}\sum_\alpha  v^\alpha_1 v^\alpha_2 f_{M}^{\alpha,(0)}
\\ & = -\frac{\partial}{\partial x_1}\left(\frac{1}{3}\rho + \rho u_1^2\right)  -\frac{\partial}{\partial x_2}\rho u_1 u_2,
\end{split}
\end{equation}
and for the second momentum derivative
\begin{equation}
\begin{split}
\Psi^{(0)}_3& = \sum_\alpha  v^\alpha_2 \left( -v^\alpha \cdot D_xf_{M}^{\alpha,(0)}\right) 
\\ & = - \frac{\partial}{\partial x_1}\sum_\alpha v^\alpha_1v^\alpha_2 f_{M}^{\alpha,(0)} - \frac{\partial}{\partial x_2}\sum_\alpha  (v^\alpha_2)^2 f_{M}^{\alpha,(0)}
\\ & = -\frac{\partial}{\partial x_1}\rho u_1 u_2  -\frac{\partial}{\partial x_2}\left( \frac{1}{3}\rho + \rho u_2^2\right).
\end{split}
\end{equation}
The first order density moment is given by,
\begin{equation}
\Psi^{(1)}_1 = \frac{1}{2}\sum_\alpha  \left( v \cdot D_x\left(v \cdot D_x f_{M}^{(0)} \right) \right) + \sum_\alpha  \left( -v \cdot  D_x f_{M}^{(1)} \right) - \frac{1}{2}\frac{\partial}{\partial t}\Psi^{(0)}_1.
\end{equation}
Again we observe that the second term is the space gradient multiplied with the momentum densities of the first order populations and hence is zero, for the first term we have
\begin{multline}
\sum_\alpha \left( v \cdot D_x\left(v \cdot D_x f_{M}^{(0)} \right) \right) \\ =  \sum_\alpha  \left( \frac{\partial^2}{\partial x_1^2} (v^\alpha_1)^2 f_{M}^{\alpha,(0)} + 2 \frac{\partial^2}{\partial x_1 \partial x_2} v^\alpha_1 v^\alpha_2 f_{M}^{\alpha,(0)}  + \frac{\partial^2}{\partial x_2^2} (v^\alpha_2)^2 f_{M}^{\alpha,(0)} \right)
  \\ = \frac{\partial^2}{\partial x_1^2}\left(\frac{1}{3}\rho + \rho u_1^2\right) + 2\frac{\partial^2}{\partial x_1 \partial x_2} \rho u_1 u_2 + \frac{\partial^2}{\partial x_2^2}\left(\frac{1}{3}\rho + \rho u_2^2\right),
\end{multline}
and for the third term
\begin{multline}
\frac{\partial}{\partial t}\Psi^{(0)}_1 =  \frac{\partial \Psi^{(0)}_1}{\partial \rho}\frac{\partial \rho}{\partial t} + \frac{\partial \Psi^{(0)}_1}{\partial \rho u_1}\frac{\partial \rho u_1}{\partial t} +\frac{\partial \Psi^{(0)}_1}{\partial \rho u_2}\frac{\partial \rho u_2}{\partial t}   = -\frac{\partial}{\partial x_1} \Psi^{(0)}_2 - \frac{\partial}{\partial x_2} \Psi^{(0)}_3
\\  = -\frac{\partial}{\partial x_1}\left(-\frac{\partial}{\partial x_1}\left(\frac{1}{3}\rho + \rho u_1^2\right)  -\frac{\partial}{\partial x_2}\rho u_1 u_2\right) \\  - \frac{\partial}{\partial x_2}\left( -\frac{\partial}{\partial x_1}\rho u_1 u_2  -\frac{\partial}{\partial x_2}\left(\frac{1}{3}\rho + \rho u_2^2\right) \right).
\end{multline}
hence subtracting these we have $ \Psi^{(1)}_1 = 0$.
\par For the first second order momentum density we have
\begin{equation}
\Psi^{(1)}_2 = \frac{1}{2}\sum_\alpha v^\alpha_1  \left( v \cdot D_x\left(v \cdot D_x f_{M}^{(0)} \right) \right) + \sum_\alpha  v^\alpha_1\left( -v \cdot  D_x f_{M}^{(1)} \right) - \frac{1}{2}\frac{\partial}{\partial t}\Psi^{(0)}_2.
\end{equation}
Examining each term in turn more closely we have for the first term
\begin{multline}
\sum_\alpha v^\alpha_1  \left( v \cdot D_x\left(v \cdot D_x f_{M}^{(0)} \right) \right) \\	 = \sum_\alpha  v^\alpha_1\left( \frac{\partial^2}{\partial x_1^2} (v^\alpha_1)^2 f_{M}^{\alpha,(0)} + 2 \frac{\partial^2}{\partial x_1 \partial x_2} v^\alpha_1 v^\alpha_2 f_{M}^{\alpha,(0)} + \frac{\partial^2}{\partial x_2^2} (v^\alpha_2)^2 f_{M}^{\alpha,(0)} \right)
\\ = \frac{\partial}{\partial x_1}\left(\rho\frac{\partial}{\partial x_1}u_1 + u_1\frac{\partial}{\partial x_1}\rho + \frac{1}{3}\rho \frac{\partial}{\partial x_2} u_2 + \frac{1}{3}u_2\frac{\partial}{\partial x_2}\rho\right) \\ + \frac{\partial}{\partial x_2}\left(\frac{1}{3}\rho\frac{\partial}{\partial x_1}u_2 + \frac{1}{3}u_2\frac{\partial}{\partial x_1}\rho + \frac{1}{3}\rho \frac{\partial}{\partial x_2} u_1 + \frac{1}{3}u_1\frac{\partial}{\partial x_2}\rho\right).
\end{multline}
The first order populations are given in Appendix \ref{app:d2q91st}, these give us for the second term
\begin{equation}
\begin{split}
&\sum_\alpha v^\alpha_1\left( -v \cdot  D_x f_{M}^{(1)} \right)  =  - \sum_\alpha v^\alpha_1\left( \frac{\partial}{\partial x_1} v^\alpha_1  f_{M}^{\alpha,(1)}  + \frac{\partial}{\partial x_2} v^\alpha_2 f_{M}^{(1)} \right) 
\\  = &\frac{\omega - 1}{\omega}\left( \frac{\partial}{\partial x_1}\left(u_1^3\frac{\partial}{\partial x_1}\rho + u_1^2u_2\frac{\partial}{\partial x_2}\rho +\left(3\rho u_1^2-\frac{2}{3}\rho\right) \frac{\partial}{\partial x_1}u_1 \right. \right. \\ & \left.  \qquad \qquad +2\rho u_1u_2\frac{\partial}{\partial x_2}u_1 + \rho u_1^2\frac{\partial}{\partial x_2}u_2  \right)  \\ &  \qquad + \frac{\partial}{\partial x_2}\left( u_1^2u_2\frac{\partial}{\partial x_1}\rho + u_1u_2^2\frac{\partial}{\partial x_2}\rho + 2\rho u_1u_2\frac{\partial}{\partial x_1} u_1 + 2\rho u_1u_2\frac{\partial}{\partial x_2} u_2\right. \\ &\left. \left. \qquad \qquad+ \left(\rho u_2^2 -\frac{1}{3}\rho\right)\frac{\partial}{\partial x_2} u_1 + \left(\rho u_1^2 -\frac{1}{3}\rho\right)\frac{\partial}{\partial x_1}u_2 \right) \right).
\end{split}
\end{equation}
and for the third term
\begin{equation}
\begin{split}
\frac{\partial}{\partial t}\Psi^{(0)}_2  &=  \frac{\partial \Psi^{(0)}_2}{\partial \rho}\frac{\partial \rho}{\partial t} + \frac{\partial \Psi^{(0)}_2}{\partial \rho u_1}\frac{\partial \rho u_1}{\partial t} +\frac{\partial \Psi^{(0)}_2}{\partial \rho u_2}\frac{\partial \rho u_2}{\partial t}  
\\  &= \left(\frac{\partial }{\partial x_1}\left( -\frac{1}{3}u^2 + u_1^2\right) +  \frac{\partial}{\partial x_2}u_1u_2 \right)\Psi^{(0)}_1  \\ & \qquad + \left( -2\frac{\partial}{\partial x_1}u_1 - \frac{\partial}{\partial x_2} u_2 \right)\Psi^{(0)}_2  - \frac{\partial}{\partial x_2} u_1 \Psi^{(0)}_3 
\\& = \frac{\partial}{\partial x_1} \left( \left(u_1 + u_1^3\right)\frac{\partial}{\partial x_1}\rho +\left(\frac{1}{3}u_2 + u_1^2u_2\right)\frac{\partial}{\partial x_2}\rho  \right. \\ & \quad \left. + \left(\frac{1}{3}\rho +3\rho u_1^2\right)\frac{\partial}{\partial x_1}u_1+ \left(\frac{1}{3}\rho +\rho u_1^2\right)\frac{\partial}{\partial x_2}u_2 + 2\rho u_1u_2\frac{\partial}{\partial x_2}u_1\right)
\\ & +\frac{\partial}{\partial x_2}\left(\left(\frac{1}{3}u_2 + u_1^2u_2\right)\frac{\partial}{\partial x_1}\rho + \left(\frac{1}{3}u_1 +u_1u_2^2\right)\frac{\partial}{\partial x_2}\rho \right. \\ & \quad \left.+ 2\rho u_1u_2\frac{\partial}{\partial x_1}u_1 +\rho u_2^2\frac{\partial}{\partial x_2}u_1 +\rho u_1^2\frac{\partial}{\partial x_1}u_2 + 2\rho u_1u_2\frac{\partial}{\partial x_2}u_2\right).
\end{split}
\end{equation}
Combining all three terms we have,
\begin{equation}
\begin{split}
\Psi_2^{(1)} &= \left(\frac{\omega - 1}{\omega} - \frac{1}{2}\right)\left( \frac{\partial}{\partial x_1}\left(u_1^3\frac{\partial}{\partial x_1}\rho + u_1^2u_2\frac{\partial}{\partial x_2}\rho +\left(3\rho u_1^2-\frac{2}{3}\rho\right) \frac{\partial}{\partial x_1}u_1 \right. \right. \\ & \left. \qquad \qquad + 2\rho u_1u_2\frac{\partial}{\partial x_2}u_1 + \rho u_1^2\frac{\partial}{\partial x_2}u_2  \right)  \\ &  \qquad + \frac{\partial}{\partial x_2}\left( u_1^2u_2\frac{\partial}{\partial x_1}\rho + u_1u_2^2\frac{\partial}{\partial x_2}\rho + 2\rho u_1u_2\frac{\partial}{\partial x_1} u_1 + 2\rho u_1u_2\frac{\partial}{\partial x_2} u_2\right. \\ &\left. \left. \qquad \qquad + \left(\rho u_2^2 -\frac{1}{3}\rho\right)\frac{\partial}{\partial x_2} u_1 + \left(\rho u_1^2 -\frac{1}{3}\rho\right)\frac{\partial}{\partial x_1}u_2 \right) \right)
\end{split}
\end{equation}The final macroscopic equations for this particular lattice and quasiequilibrium then are to first order
\begin{equation}
\begin{split}
& \frac{\partial}{\partial t}\rho  = -D_x \rho u \\ 
& \frac{\partial}{\partial t}\rho u_1  = -\frac{\partial}{\partial x_1}\left(\frac{\rho}{3} +\rho u_1^2\right)-\frac{\partial}{\partial x_2}\rho u_1u_2 \\
& \qquad - \epsilon \frac{2-\omega}{2\omega}\left( \frac{\partial}{\partial x_1}\left(u_1^3\frac{\partial}{\partial x_1}\rho + u_1^2u_2\frac{\partial}{\partial x_2}\rho +\left(3\rho u_1^2-\frac{2}{3}\rho\right) \frac{\partial}{\partial x_1}u_1 \right. \right. \\  &\qquad \qquad \qquad \left. + 2\rho u_1u_2\frac{\partial}{\partial x_2}u_1 + \rho u_1^2\frac{\partial}{\partial x_2}u_2  \right)  \\ & \qquad \qquad + \frac{\partial}{\partial x_2}\left( u_1^2u_2\frac{\partial}{\partial x_1}\rho + u_1u_2^2\frac{\partial}{\partial x_2}\rho + 2\rho u_1u_2\frac{\partial}{\partial x_1} u_1 + 2\rho u_1u_2\frac{\partial}{\partial x_2} u_2\right. \\ &\left. \left. \qquad \qquad \qquad+ \left(\rho u_2^2 -\frac{1}{3}\rho\right)\frac{\partial}{\partial x_2} u_1 + \left(\rho u_1^2 -\frac{1}{3}\rho\right)\frac{\partial}{\partial x_1}u_2 \right) \right)
\end{split}
\end{equation}
The second momentum density is available easily through symmetry. 

\section{Continuous Velocity Examples}
We are now concerned with calculating, via the invariant manifold populations, the macroscopic moments approximated by
the LBM chain in a continuous velocity system. We select two examples, chosen to match the previous discrete velocity
schemes.
\subsection{The athermal 1-D model}
The first continuous velocity model we will examine is one chosen to match the zero order dynamics of the discrete model studied in section \ref{sec:athermal3vel1d}, the one dimensional system with the three discrete velocities $\{-1,0,1\}$. The continuous population function acting as the quasi-equilibrium is a specific case of the Maxwell distribution where the temperature is fixed, in this case to $1/3$.
\begin{equation}
f^{(0)} = \rho \sqrt{\frac{3}{2\pi}}\exp\left(-\frac{3}{2}(v-u)^2\right)
\end{equation}
With such a system the macroscopic variables are calculated as integrals rather than the sums in the discrete case.
\begin{equation}
\begin{split}
\int_{-\infty}^{\infty}f^{(0)}dv & = \rho \\
\int_{-\infty}^{\infty}vf^{(0)}dv & = \rho u \\
\int_{-\infty}^{\infty}v^2f^{(0)}dv & = \frac{1}{3}\rho  + \rho u^2 
\end{split}
\end{equation}
Clearly this matches the moments retrieved in the discrete velocity case. Due to this we can, analagously to the discrete case, immediately write down the zero order macroscopic dynamics following equation \ref{eq:psi0eq}.
\begin{equation}
\Psi_0^{(1)} = -\int_\infty^{\infty}v\left(\frac{\partial}{\partial x}f^{(0)}\right)dv = -\frac{\partial}{\partial x}\int_\infty^{\infty}vf^{(0)}dv = - \frac{\partial}{\partial x} \rho u 
\end{equation}
\begin{equation}
\Psi_0^{(2)} = -\int_\infty^{\infty}v^2\left(\frac{\partial}{\partial x}f^{(0)}\right)dv = -\frac{\partial}{\partial x}\int_\infty^{\infty}v^2f^{(0)}dv = - \frac{\partial}{\partial x} \left(\frac{1}{3}\rho  + \rho u^2\right)
\end{equation}
In order to calculate the first order moments we expect that we shall require the first order continuous populations. These are also derived exactly as in the discrete case with the replacement of the sum, by the integral, in the calculation of the moments. Since we replicate the discrete case the collision we select is again the BGK collision and we derive the first order populations from equation \ref{eq:f1bgk}.
\begin{equation}
\frac{\omega}{1-\omega}f^{(1)} = \rho\sqrt{\frac{3}{2\pi}}\exp\left(-\frac{3}{2}(v-u)^2\right)\cdot\left(1-3v^2+6vu-3u^2\right)\cdot\left(\frac{\partial}{\partial x} u\right)
\end{equation}
Again we calculate the first order moments from the template given by equation \ref{eq:psi1eq}
\begin{equation}
\Psi^{(1)}_1 =  \frac{1}{2} \int_\infty^\infty v^2 \left(\frac{\partial^2}{\partial x^2}f^{(0)}\right)dv  - \int_\infty^\infty v \left(\frac{\partial}{\partial x}f^{(1)}\right)dv - \frac{1}{2}\frac{\partial}{\partial t}\Psi^{(0)}_1
\end{equation}
Exactly as the discrete case the second term here is the space derivative of the momentum of the $f^{(1)}$ which equals zero due to all moments of non equilibrium components being zero and the first term can be calculated immediately from the quasi-equilibrium therefore,
\begin{equation}
\Psi^{(1)}_1 =   \frac{1}{2}\frac{\partial^2}{\partial x^2} \left(\frac{\rho}{3} + \rho u^2\right)  - \frac{1}{2}\frac{\partial}{\partial t}\Psi^{(0)}_1.
\end{equation}
Again the time derivative of $\Psi^{(0)}$ can be calculated by the chain rule,
\begin{equation}
\frac{\partial}{\partial t}\Psi^{(0)}_1 = \frac{\partial \Psi^{(0)}_1}{\partial \rho}\frac{\partial \rho}{\partial t} + \frac{\partial \Psi^{(0)}_1}{\partial \rho u}\frac{\partial \rho u}{\partial t}  = -\frac{\partial}{\partial x} \Psi^{(0)}_2 =  \frac{\partial^2}{\partial x^2} \left(\frac{\rho}{3} + \rho u^2\right)
\end{equation}
Substituting we have,
\begin{equation}
\Psi^{(1)}_1 =  \frac{1}{2}\frac{\partial^2}{\partial x^2} \left(\frac{\rho}{3} + \rho u^2\right) - \frac{1}{2}\frac{\partial^2}{\partial x^2} \left(\frac{\rho}{3} + \rho u^2\right) = 0 .
\end{equation}
For the continuous velocity momentum moment we have
\begin{equation}
\Psi^{(1)}_2 =  \frac{1}{2} \int_\infty^\infty v^3 \left(\frac{\partial^2}{\partial x^2}f{(0)}\right)dv - \int_\infty^\infty v^2 \left(\frac{\partial}{\partial x}f^{(1)}\right)dv - \frac{1}{2}\frac{\partial}{\partial t}\Psi^{(0)}_2
\end{equation}
Rearranging and performing the first two integrals gives us
\begin{equation}
\Psi^{(1)}_2 =  \frac{1}{2}\frac{\partial^2}{\partial x^2}\left(\rho u + \rho u^3\right) - \frac{\partial}{\partial x}\left(-\frac{2}{3}\rho\left(\frac{\partial}{\partial x} u\right)\right) - \frac{1}{2}\frac{\partial}{\partial t}\Psi^{(0)}_2
\end{equation}
Again using the chain rule, this term is exactly as in the discrete case,
\begin{multline}
\frac{\partial}{\partial t}\Psi^{(0)}_2 = \frac{\partial \Psi^{(0)}_2}{\partial \rho}\frac{\partial \rho}{\partial t} + \frac{\partial \Psi^{(0)}_2}{\partial \rho u}\frac{\partial \rho u}{\partial t}  = -\frac{\partial}{\partial x}\left(\frac{1}{3} - u^2\right)\Psi^{(0)}_1 - \frac{\partial}{\partial x} 2u \Psi^{(0)}_2 \\ = \frac{\partial}{\partial x}\left(\frac{2}{3}u\frac{\partial}{\partial x}\rho +\left(\frac{1}{3}- u^2\right)\frac{\partial}{\partial x}\rho u  + 2u\frac{\partial}{\partial x}\rho u^2\right)
\end{multline}
Substituting this all back in we have,
\begin{equation}
\Psi^{(1)}_2 =  \left(\frac{\omega-1}{\omega} -\frac{1}{2}\right)\frac{\partial}{\partial x}\left( -\frac{2}{3}\rho\frac{\partial}{\partial x}u \right)= \frac{\omega - 2}{2\omega}\frac{\partial}{\partial x}\left( -\frac{2}{3}\rho\frac{\partial}{\partial x}u \right) .
\end{equation}	
The moment gradients are then, for the continuous velocity system, to first order in $\epsilon$,
\begin{equation}
\begin{split}
&\frac{\partial}{\partial t}\rho = -\frac{\partial}{\partial x} \rho u + o(\epsilon)\\
&\frac{\partial}{\partial t}\rho u = -\frac{\partial}{\partial x}\left(\frac{1}{3}\rho + \rho u^2\right)-\epsilon\frac{2-\omega}{2\omega}\frac{\partial}{\partial x}\left( -\frac{2}{3}\rho\frac{\partial}{\partial x}u\right)+o(\epsilon)
\end{split}
\end{equation}
We immediately observe that several of the dissipative terms that appeared in the discrete velocity system do not occur when we use continuous velocities
\subsection{The athermal 2D model}
The next continuous velocity model we examine is the widely used athermal 2d model. Again we use a specific choice of the Maxwellian distribution which matches the zero order moments given by the discrete velocity set.
\begin{equation}
f^{(0)} = \rho\frac{3}{2\pi} \exp\left(-\frac{3}{2}\left((v_1-u_1)^2 + (v_2-u_2)^2\right)\right)
\end{equation}
Again macroscopic variables are calculated by integrals over velocity space
\begin{equation}
\begin{split}
\int_{\mathbb{R}^2} f^{(0)}dv & = \rho \\
\int_{\mathbb{R}^2} v_1f^{(0)}dv & = \rho u_1 \\
\int_{\mathbb{R}^2} v_2f^{(0)}dv & = \rho u_2 \\
\int_{\mathbb{R}^2} (v_1^2+v_2^2)f^{(0)}dv & = \frac{2}{3}\rho  + \rho (u_1^2 +u_2^2)
\end{split}
\end{equation}
Again we calculate the zero order moments,
\begin{equation}
\begin{split}
\Psi^{(0)}_1& = \int_{\mathbb{R}^2}  -v \cdot  D_xf_{M}^{\alpha,(0)} dv
\\ & = - \frac{\partial}{\partial x_1}\int_{\mathbb{R}^2}  v_1 f_{M}^{\alpha,(0)}dv - \frac{\partial}{\partial x_2}\int_{\mathbb{R}^2}  v_2 f_{M}^{\alpha,(0)}dv
\\ & = -\frac{\partial}{\partial x_1} \rho u_1 - \frac{\partial}{\partial x_2} \rho u_2
 \end{split}
\end{equation}
and for the first momentum derivative
\begin{equation}
\begin{split}
\Psi^{(0)}_2& = \int_{\mathbb{R}^2}  v_1 \left( -v \cdot D_xf_{M}^{\alpha,(0)}\right) dv 
\\ & = - \frac{\partial}{\partial x_1}\int_{\mathbb{R}^2}  v_1^2 f_{M}^{\alpha,(0)}dv - \frac{\partial}{\partial x_2}\int_{\mathbb{R}^2}  v_1 v_2 f_{M}^{\alpha,(0)}dv
\\ & = -\frac{\partial}{\partial x_1}\left(\frac{1}{3}\rho + \rho u_1^2\right)  -\frac{\partial}{\partial x_2}\rho u_1 u_2
\end{split}
\end{equation}
for the second momentum derivative
\begin{equation}
\begin{split}
\Psi^{(0)}_3& = \int_{\mathbb{R}^2}  v_2 \left( -v \cdot D_xf_{M}^{\alpha,(0)}\right)dv 
\\ & = - \frac{\partial}{\partial x_1}\int_{\mathbb{R}^2} v_1v_2 f_{M}^{\alpha,(0)}dv - \frac{\partial}{\partial x_2}\int_{\mathbb{R}^2}  v_2^2 f_{M}^{\alpha,(0)}dv
\\ & = -\frac{\partial}{\partial x_1}\rho u_1 u_2  -\frac{\partial}{\partial x_2}\left( \frac{1}{3}\rho + \rho u_2^2\right)
\end{split}
\end{equation}
We again calculate the first order populations following equation \ref{eq:f1bgk}.
\begin{equation}
\begin{split}
&\frac{\omega}{1-\omega}f^{(1)}=
\\ &  \quad \rho\frac{3}{2\pi} \exp\left(-\frac{3}{2}\left((v_1-u_1)^2 + (v_2-u_2)^2\right)\right)
\\ &  \qquad \cdot\left(\left(1-3v_1^2+6v_1u_1-3u_1^2\right)\frac{\partial}{\partial x_1}u_1 \right. 
\\ & \qquad \qquad + \left(-3v_1v_2+3v_1u_2+3v_2u_1-3u_1u_2\right)\frac{\partial}{\partial x_2}u_1
\\ & \qquad \qquad + \left(-3v_1v_2+3v_1u_2+3v_2u_1-3u_1u_2\right)\frac{\partial}{\partial x_1}u_2 
\\ &\left. \qquad \qquad +  \left(1-3v_2^2+6v_2u_2-3u_2^2\right)\frac{\partial}{\partial x_2}u_2\right)
\end{split}
\end{equation}
The first order density moment is given by,
\begin{equation}
\Psi^{(1)}_1 = \frac{1}{2} \int_{\mathbb{R}^2}  v \cdot D_x\left(v \cdot D_x f^{(0)} \right) dv - \int_{\mathbb{R}^2}  v \cdot  D_x f_{M}^{(1)}dv - \frac{1}{2}\frac{\partial}{\partial t}\Psi^{(0)}_1
\end{equation}
Performing the integrals of the first two terms we note that the second term is again zero therefore
\begin{equation}
\Psi^{(1)}_1 = \frac{1}{2}\frac{\partial^2}{\partial x_1^2}\left(\frac{1}{3}\rho + \rho u_1^2\right)  + \frac{\partial^2}{\partial x_1x_2}\rho u_1 u_2 +\frac{1}{2} \frac{\partial^2}{\partial x_2^2}\left(\frac{1}{3}\rho + \rho u_2^2\right)  - \frac{1}{2}\frac{\partial}{\partial t}\Psi^{(0)}_1\end{equation}
and exactly as in the discrete velocity system we have for the third term
\begin{equation}
\begin{split}
\frac{\partial}{\partial t}\Psi^{(0)}_1 = & \frac{\partial \Psi^{(0)}_1}{\partial \rho}\frac{\partial \rho}{\partial t} + \frac{\partial \Psi^{(0)}_1}{\partial \rho u_1}\frac{\partial \rho u_1}{\partial t} +\frac{\partial \Psi^{(0)}_1}{\partial \rho u_2}\frac{\partial \rho u_2}{\partial t}   = -\frac{\partial}{\partial x_1} \Psi^{(0)}_2 - \frac{\partial}{\partial x_2} \Psi^{(0)}_3
\\  = & -\frac{\partial}{\partial x_1}\left(-\frac{\partial}{\partial x_1}\left(\frac{1}{3}\rho + \rho u_1^2\right)  -\frac{\partial}{\partial x_2}\rho u_1 u_2\right) 
\\ & \qquad - \frac{\partial}{\partial x_2}\left( -\frac{\partial}{\partial x_1}\rho u_1 u_2  -\frac{\partial}{\partial x_2}\left(\frac{1}{3}\rho + \rho u_2^2\right) \right)
\end{split}
\end{equation}
hence subtracting these we have $ \Psi^{(1)}_1 = 0$.
\par For the first second order momentum density we have
\begin{equation}
\Psi^{(1)}_2 = \frac{1}{2}\int_{\mathbb{R}^2} v_1  \left( v \cdot D_x\left(v \cdot D_x f^{(0)} \right) \right)dv + \int_{\mathbb{R}^2}  v_1\left( -v \cdot  D_x f^{(1)} \right)dv - \frac{1}{2}\frac{\partial}{\partial t}\Psi^{(0)}_2
\end{equation}
Again performing the integrations from the first two terms we have
\begin{equation}
\begin{split}
\Psi^{(1)}_2 = & \frac{1}{2}\left(\frac{\partial^2}{\partial x_1^2}\left(\rho u_1^3  +\rho u_1\right) + \frac{\partial^2}{\partial x_1 \partial x_2}\left(\frac{1}{3}\rho u_2 +\rho u_1^2u_2\right) \right. \\ & \qquad \left. + \frac{\partial^2}{\partial x_2}\left(\frac{1}{3}\rho u_1 + \rho u_1u_2^2\right)\right) \\ & +\frac{\omega-1}{2\omega}\left( \frac{\partial}{\partial x_1}\left(-\frac{2}{3}\rho\frac{\partial}{\partial x_1}u_1\right) \right. \\ & \qquad \left. - \frac{\partial}{\partial x_2}\left(-\frac{1}{3}\rho\left(\frac{\partial}{\partial x_1}u_2 + \frac{\partial}{\partial x_2}u_1\right)\right)\right)- \frac{1}{2}\frac{\partial}{\partial t}\Psi^{(0)}_2
\end{split}
\end{equation}
and for the third term
\begin{equation}
\begin{split}
\frac{\partial}{\partial t}\Psi^{(0)}_2  &=  \frac{\partial \Psi^{(0)}_2}{\partial \rho}\frac{\partial \rho}{\partial t} + \frac{\partial \Psi^{(0)}_2}{\partial \rho u_1}\frac{\partial \rho u_1}{\partial t} +\frac{\partial \Psi^{(0)}_2}{\partial \rho u_2}\frac{\partial \rho u_2}{\partial t}  
\\  &= \left(\frac{\partial }{\partial x_1}\left( -\frac{1}{3}u^2 + u_1^2\right) +  \frac{\partial}{\partial x_2}u_1u_2 \right)\Psi^{(0)}_1 
\\ & \qquad + \left( -2\frac{\partial}{\partial x_1}u_1 - \frac{\partial}{\partial x_2} u_2 \right)\Psi^{(0)}_2  - \frac{\partial}{\partial x_2} u_1 \Psi^{(0)}_3 
\\& = \frac{\partial}{\partial x_1} \left( \left(u_1 + u_1^3\right)\frac{\partial}{\partial x_1}\rho +\left(\frac{1}{3}u_2 + u_1^2u_2\right)\frac{\partial}{\partial x_2}\rho \right.
\\& \qquad \left. + \left(\frac{1}{3}\rho +3\rho u_1^2\right)\frac{\partial}{\partial x_1}u_1 + \left(\frac{1}{3}\rho +\rho u_1^2\right)\frac{\partial}{\partial x_2}u_2 + 2\rho u_1u_2\frac{\partial}{\partial x_2}u_1\right)
\\ & +\frac{\partial}{\partial x_2}\left(\left(\frac{1}{3}u_2 + u_1^2u_2\right)\frac{\partial}{\partial x_1}\rho + \left(\frac{1}{3}u_1 +u_1u_2^2\right)\frac{\partial}{\partial x_2}\rho + 2\rho u_1u_2\frac{\partial}{\partial x_1}u_1 \right.
\\& \qquad + \left. \rho u_2^2\frac{\partial}{\partial x_2}u_1 +\rho u_1^2\frac{\partial}{\partial x_1}u_2 + 2\rho u_1u_2\frac{\partial}{\partial x_2}u_2\right)
\end{split}
\end{equation}
Combining all three terms we have
\begin{multline}
\Psi_2^{(1)} = \left(\frac{\omega - 1}{\omega} - \frac{1}{2}\right)\left(\frac{\partial}{\partial x_1}\left(-\frac{2}{3}\rho\frac{\partial}{\partial x_1}u_1\right) \right. \\ \left. - \frac{\partial}{\partial x_2}\left(-\frac{1}{3}\rho\left(\frac{\partial}{\partial x_1}u_2 + \frac{\partial}{\partial x_2}u_1\right)\right) \right)
\end{multline}
The final macroscopic equations for this particular lattice and quasiequilibrium then are 
\begin{equation}
\begin{split}
& \frac{\partial}{\partial t}\rho  = -D_x \rho u \\ 
& \frac{\partial}{\partial t}\rho u_1  = -\frac{\partial}{\partial x_1}\left(\frac{\rho}{3} +\rho u_1^2\right)-\frac{\partial}{\partial x_2}\rho u_1u_2 \\
& \qquad - \epsilon \frac{2-\omega}{2\omega}\left(\frac{\partial}{\partial x_1}\left(-\frac{2}{3}\rho\frac{\partial}{\partial x_1}u_1\right) - \frac{\partial}{\partial x_2}\left(-\frac{1}{3}\rho\left(\frac{\partial}{\partial x_1}u_2 + \frac{\partial}{\partial x_2}u_1\right)\right) \right)
\end{split}
\end{equation}
and again the second momentum density can be found by reflection. Once again many of the disippative terms vanish in the continuous velocity system.

\section{Macroscopic Stability}
\index{Macroscopic stability}In the previous sections we have demonstrated the discrete velocity systems studied do not recover the exact macroscopic
dissipative dynamics of the continuous system. We are now concerned with the stability of the discrete dynamics under a
short wave perturbation. In each example we are concerned with the stability of the linear part of the dynamics (as calculated above) only.

\subsection{The athermal 1-D model}
We consider perturbations by a Fourier mode around a constant flow, that is we write
\begin{equation}
\begin{split}
&\rho = \rho_0 + Ae^{i(\lambda t + \kappa x)} \\
&u = u_0 + Be^{i(\lambda t + \kappa x)}
\end{split}
\end{equation}
We combine this with a composite coefficient for the first order part
\begin{equation}
\nu = \epsilon\frac{2-\omega}{2\omega}
\end{equation}
Substituting these into the macroscopic equations and with some rearrangement for the $u$ term we have
\begin{equation}
\begin{split}
&A\lambda = -\rho_0B\kappa-u_0A\kappa \\
&B\lambda = -\frac{1}{3\rho_0}A\kappa -u_0B\kappa - \nu\left(\frac{u_0^3}{\rho_0}Ai\kappa^2+\left(3u_0^2-\frac{2}{3}\right)Bi\kappa^2\right)
\end{split}
\end{equation}
We take eigenvalues of the matrix
\begin{equation}
\left(\begin{array}{cc}
-u_0\kappa & -\rho_0\kappa \\
-\frac{1}{3\rho_0}\kappa-\nu\frac{u_0^3}{\rho_0}i\kappa^2 &-u_0\kappa-\nu\left(3u_0^2-\frac{2}{3}\right)i\kappa^2
\end{array}\right)
\end{equation}
which give us two values for $\lambda$
\begin{multline}
\lambda = \kappa\left(-u_0-\frac{3}{2}\nu u_0^2i\kappa+\frac{1}{3}\nu i\kappa \right. \\ \left. \pm \sqrt{\nu u_0^3i\kappa - \frac{9}{4}\nu^2u_0^4\kappa^2+\nu^2u_0^2\kappa^2-\frac{1}{9}\nu^2\kappa^2 +\frac{1}{3}}\right)
\end{multline}
In order for the manifold to remain bounded in time we investigate parameters which give $\Im(\lambda) \geq 0$. We begin by checking aymptotics of two parameters, for large $\kappa$ we have
\begin{equation}
\begin{split}
\lambda &=\nu\kappa^2\left(-\frac{3}{2}u_0^2i+\frac{1}{3}i\pm\sqrt{-\left(\frac{3}{2}u_0^2+\frac{1}{3}\right)}\right) \\
& =0,\nu\kappa^2\left(-3u_0^2+\frac{2}{3}\right)i
\end{split}
\end{equation}
and for large $u_0$
\begin{equation}
\begin{split}
\lambda & =\kappa\left(-\frac{3}{2}\nu u_0^2i\kappa \pm \sqrt{-\frac{9}{4}\nu^2 u_0^4 \kappa^2}\right) \\
&=0,-3\nu u_0^2i\kappa^2
\end{split}
\end{equation}
We can see from this that for non-zero $\kappa$ the first condition that should be satisfied for stability is $u_0^2<2/9$, for large $u_0$ it is necessary for $\kappa$ to equal 0. Additionally, stability is absolutely contingent on the composite coefficient $\nu$ being positive, this is the dual condition that time steps are positive and that relaxation parameter of the collision $\omega$ is in the interval $0\leq\omega\leq2$ (repeated steps of the collision integral in isolation go towards the quasiequilibrium). In the case that either $\nu$ is negative or that $u_0$ is outside the given region, the magnitude of the Fourier perturbation will grow exponentially causing a rapid divergence from the constant flow. 
\par
\begin{figure}[h]
\centering
\includegraphics[width=\textwidth]{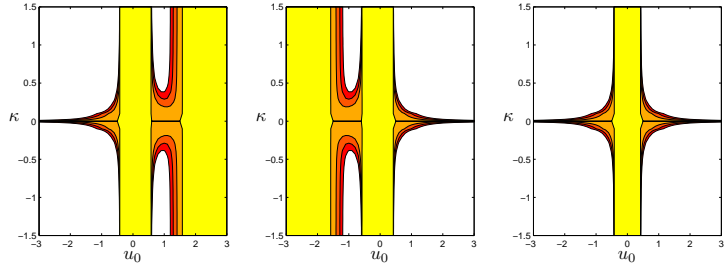}
\caption{The first two figures show stability for each of the two eigenvalues in the D1Q3 system with $\nu = 1$, the third figure plots the minimum of the two. Contours are plotted at $\Im (\lambda) = (-0.3,-0.2,-0.1,0)$, the yellow region and it's boundary indicate the stable region and, the other colours, the decay from stability.}
\label{fig:d1q3stab}
\end{figure}

 We can confirm these results numerically by plotting the contours of the two eigenvalues equal to zero. In fact in Figure(\ref{fig:d1q3stab}) we additionally plot contours below zero to show the decay from stability.

\subsection{The athermal 2-D model}
\begin{figure}[h!]
\centering
\includegraphics[width=0.5\textwidth]{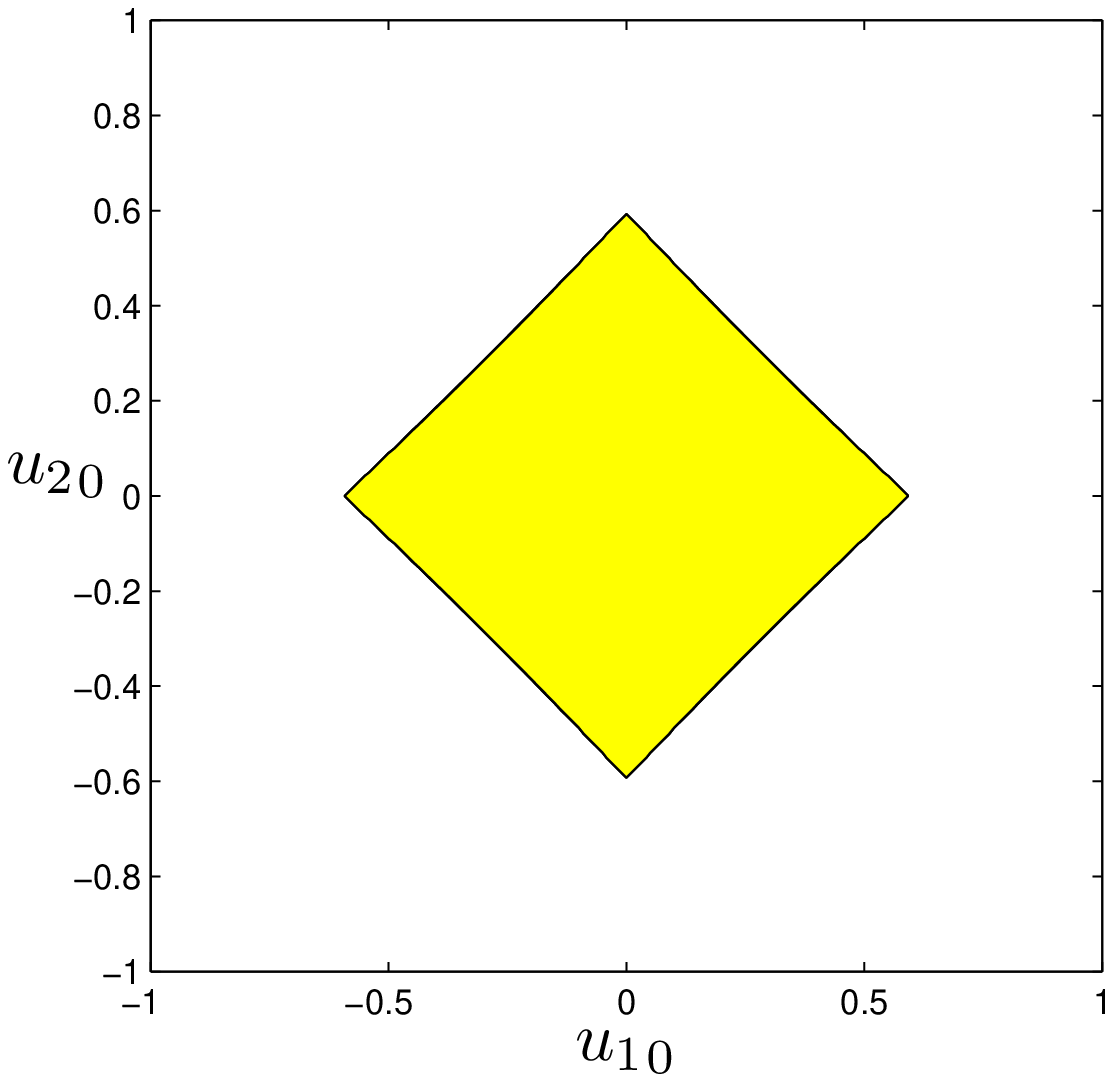}
\caption{Stability regions for the athermal D2Q9 system. Contours are plotted at $\Im (\lambda) = (-0.3,-0.2,-0.1,0)$, the yellow region and it's boundary indicate the stable region and, the other colours, the decay from stability. The parameters are $\nu = 1$ and  $|\kappa_1| ,|\kappa_2| \rightarrow \infty$.}
\label{fig:d2q9stab4}
\end{figure}

We extend the stability analysis from the one dimensional case with a perturbation in the additional space direction. The perturbed system is given by,
\begin{equation}
\begin{split}
&\rho = \rho_0 + Ae^{i(\lambda t + \kappa_1 x_1 + \kappa_2 x_2)}\\
&u_1 = {u_1}_0 + B_1e^{i(\lambda t + \kappa_1 x_1 + \kappa_2 x_2)}\\
&u_2 = {u_2}_0 + B_2e^{i(\lambda t + \kappa_1 x_1 + \kappa_2 x_2)}
\end{split}
\end{equation}
In this case we investigate the short wave asymptotics as $|\kappa_1|,|\kappa_2|\rightarrow \infty$. The eigenvalues of the system under such conditions are
\begin{equation}
\begin{split}
\lambda_{1,2} = & \left(\frac{1}{3}-\frac{3}{2}{u_1}_0^2\right)i\nu\kappa_1^2 + \left(\frac{1}{3}-\frac{3}{2}{u_2}_0^2\right)i\nu\kappa_2^2 - 3i\nu {u_1}_0{u_2}_0\kappa_1\kappa_2 \\ & \qquad \pm \sqrt{-\left(\left(\frac{1}{3}-\frac{3}{2}{u_1}_0^2\right)\nu\kappa_1^2 + \left(\frac{1}{3}-\frac{3}{2}{u_2}_0^2\right)\nu\kappa_2^2 - 3\nu {u_1}_0{u_2}_0\kappa_1\kappa_2\right)}, 
\\ \lambda_3 =& \left(\frac{1}{3}-{u_1}_0^2\right)i\nu\kappa_1^2 + \left(\frac{1}{3}-{u_2}_0^2\right)i\nu\kappa_2^2 - 2i\nu {u_1}_0{u_2}_0\kappa_1\kappa_2.
\end{split}
\end{equation}
In the 1-D examples all terms were in even powers of $\kappa$ whereas in this case there are cross terms in the product $\kappa_1\kappa_2$. Because of this it is necessary to consider the different permutations of signs for these terms. Since the condition that the third eigenvalue imposes is weaker than that of the the first two, which are equivalent, it is sufficient to find the region of stability using just one of these. Again assuming that the coefficient $\nu$ is positive, the region is given by parameters satisfying the two conditions.
\begin{equation}
\begin{split}
\left(\frac{1}{3}-\frac{3}{2}{u_1}_0^2\right) + \left(\frac{1}{3}-\frac{3}{2}{u_2}_0^2\right)- 3{u_1}_0{u_2}_0 \geq 0 \\
\left(\frac{1}{3}-\frac{3}{2}{u_1}_0^2\right) + \left(\frac{1}{3}-\frac{3}{2}{u_2}_0^2\right)+ 3{u_1}_0{u_2}_0 \geq 0
\end{split}
\end{equation}
The plot of the region generated by these inequalities is given in Figure \ref{fig:d2q9stab4}. Similarly to the one dimensional examaple, in the event that $\nu$ is negative or the constant flow speed moves outside this region, the magnitude of the Fourier perturbation will increase exponentially in time.
\par
\begin{figure}[h!]
\centering
\includegraphics[width=\textwidth]{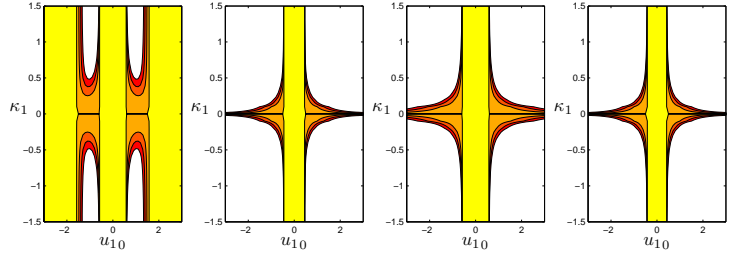}
\caption{The first three figures show stability for each of the three eigenvalues in the athermal D2Q9 system with parameters $\nu =1,u_2 = 0, \kappa_2=0$, the fourth figure plots the minimum of them. In each case the contours are plotted at $\lambda = \left(-0.3,-0.2,-0.1,0\right)$ therefore the yellow region and it's boundary describe the stable area.}
\label{fig:d2q9stab1}
\end{figure}

\begin{figure}[h!]
\centering
\includegraphics[width=\textwidth]{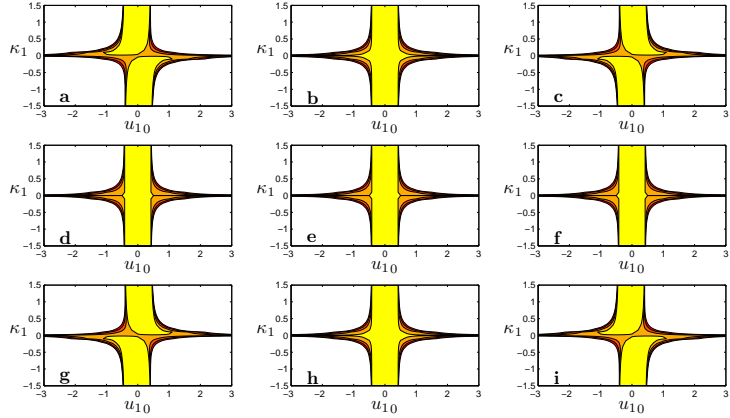}
\caption{Stability regions for the athermal D2Q9 system. Contours are plotted at $\Im (\lambda) = (-0.3,-0.2,-0.1,0)$, the yellow region and it's boundary indicate the stable region and, the other colours, the decay from stability. The parameters are $\nu = 1$ and additionally \textbf{a)} $\kappa_2 = -0.1,{u_2}_0 = -0.5$;\textbf{b)} $\kappa_2 = -0.1,{u_2}_0 = 0$;\textbf{c)} $\kappa_2 = -0.1,{u_2}_0 = 0.5$; \textbf{d)} $\kappa_2 = 0,{u_2}_0 = -0.5$; \textbf{e)} $\kappa_2 = 0,{u_2}_0 = 0$; \textbf{f)} $\kappa_2 = 0,{u_2}_0 = 0.5$; \textbf{g)} $\kappa_2 = 0.1,{u_2}_0 = -0.5$;\textbf{h)} $\kappa_2 = 0.1,{u_2}_0 = 0$;\textbf{i)} $\kappa_2 = 0.1,{u_2}_0 = 0.5$.}
\label{fig:d2q9stab2}
\end{figure}

\begin{figure}[h!]
\centering
\includegraphics[width=\textwidth]{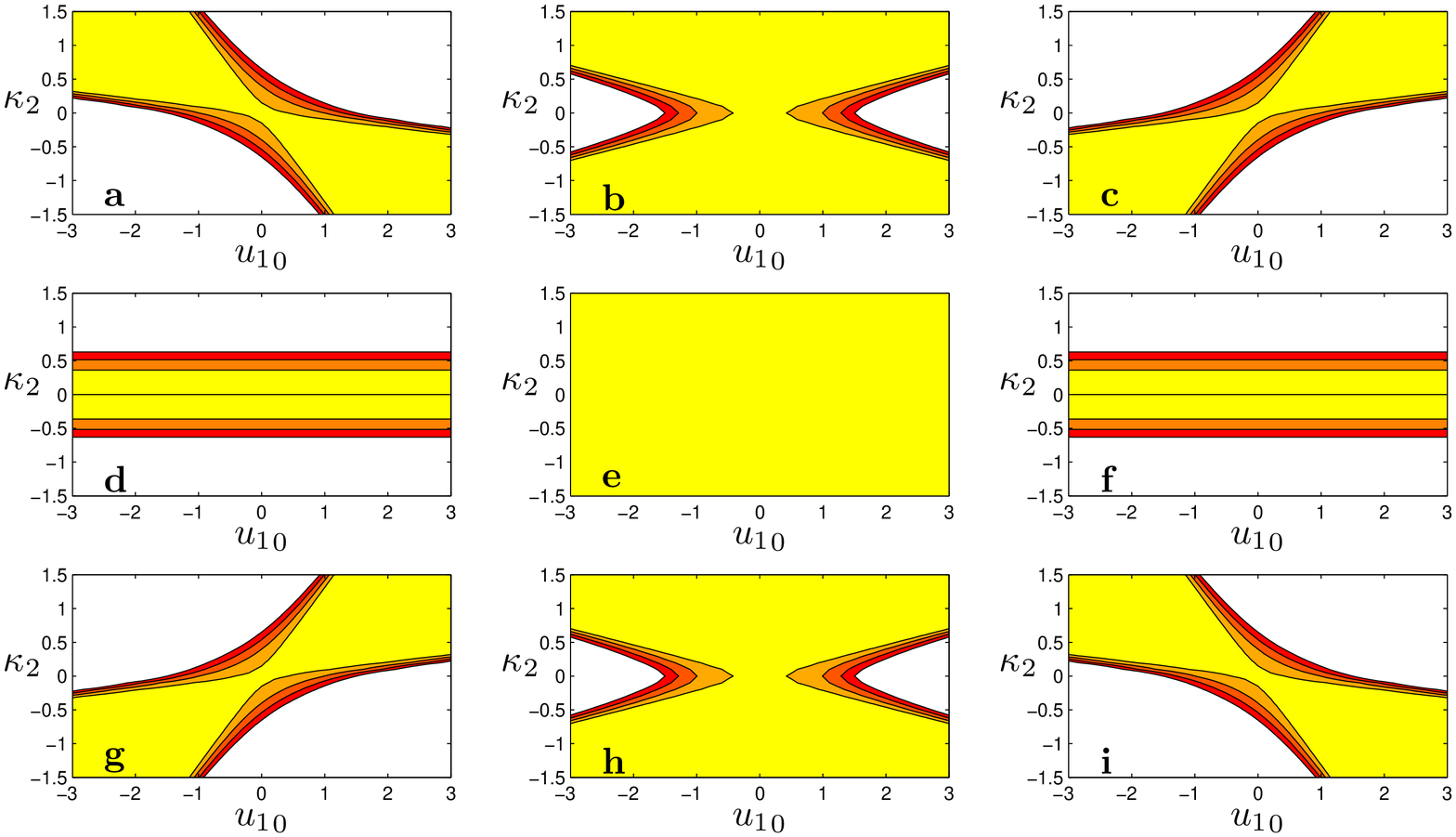}
\caption{Stability regions for the athermal D2Q9 system. Contours are plotted at $\Im (\lambda) = (-0.3,-0.2,-0.1,0)$, the yellow region and it's boundary indicate the stable region and, the other colours, the decay from stability. The parameters are $\nu = 1$ and additionally \textbf{a)} $\kappa_1 = -0.1,{u_2}_0 = -0.5$;\textbf{b)} $\kappa_1 = -0.1,{u_2}_0 = 0$;\textbf{c)} $\kappa_1 = -0.1,{u_2}_0 = 0.5$; \textbf{d)} $\kappa_1 = 0,{u_2}_0 = -0.5$; \textbf{e)} $\kappa_1 = 0,{u_2}_0 = 0$; \textbf{f)} $\kappa_1 = 0,{u_2}_0 = 0.5$; \textbf{g)} $\kappa_1 = 0.1,{u_2}_0 = -0.5$;\textbf{h)} $\kappa_1 = 0.1,{u_2}_0 = 0$;\textbf{i)} $\kappa_1 = 0.1,{u_2}_0 = 0.5$.}
\label{fig:d2q9stab3}
\end{figure}

 Again for specific parameters the stability can be calculated numerically. In the first case examine the case where $\kappa_2,{u_2}_0 = 0$. Figure (\ref{fig:d2q9stab1}) shows the stability plot for the three eigenvalues and their minimum, in this case we see that while the eigenvalues are different from their counterparts in the 1-D system, the stability region is exactly the same.
In Figure (\ref{fig:d2q9stab2}) we vary $\kappa_2$ and ${u_2}_0$ to see what affect this has on the stability region.
For a more complete picture we plot ${u_1}_0$ against $\kappa_2$ and again plot the stability region. In Figure (\ref{fig:d2q9stab3}) we vary $\kappa_1$ and ${u_2}_0$ across the different plots.

\section{Conclusion}
In the analaysis of the continuous Boltzmann equation, the Chapman-Enskog procedure is known to reproduce the Navier Stokes equations\cite{succitlbe,karlince,luoce}. This is achieved by a perturbation by a small parameter, the Knudsen number. At the zero and first orders of this parameter, repsectively, the convective and diffusive dynamics appear. At higher orders which were not discussed here the Burnett equations arise.
\par The discrete Boltzmann schemes studied here are defined by the requirement that the Euler equations are recovered at the zero order. In common with the continuous scheme, dissipative terms arise at the first order, however in the discrete case there appear additional viscous terms. In parallel with Goodman and Lax\cite{laxdispersive} we view the additional dissipative part of the fluid as a direct consequence of the discrete scheme used. In this work we have used the idea of invariant manifolds\cite{invariantmanifolds} to calculate the macroscopic dynamics arising from discrete time Boltzmann schemes. This technique is based on an expansion in a different small parameter, the time step $\epsilon$. Dynamics at the zero and first orders again correspond to the conservative and dissipative parts of a fluid respectively. Although in this work we calculate these dynamics up to the first order only, the methodology can be extended to calculate higher order systems.
\par To compute a solution to the Boltzmann system it is also necessary to discretize velocity space. We have presented two alternative modes of thought to reason why this should be, which produce equivalent systems. We have calculated the exact macroscopic dynamics up to first order of two common discrete velocity schemes, and their
continuous counterparts. Although the dynamics of these two schemes match at the zero order, in the discrete velocity case
additional erroneous terms arise at the first order. Such errors might be expected due to the way the quasi-equilibria in
the discrete case are defined. If we view the discrete velocity summation as a quadrature approximation to the continuous
velocity integral, then we should expect an error of integration. At the zero order we find no such error. This is due to an
equilibrium being constructed specifically that the zero order moments are calculated exactly. This equilibrium consists of
merely the first three terms of the Taylor expansion of the continuous equilibrium about the zero momentum position. It
should, perhaps then, be no surprise that the dissipative dynamics in the discrete system approach those of the continuous
system only in the limit of momentum going to zero.
\par Finally we perform a stability analysis of the linear part of the macroscopic dynamics of the discrete velocity schemes under a short wave perturbation. In common with other authors using similar Fourier techniques \cite{sterlingstab,servanstab}, and with our own earlier assumptions, we find that two lattice parameters are critical for stability. These are the time step $\epsilon$ which must be positive, and the relaxation parameter $\omega$, which must be chosen for non-zero flow speed in the interval $(-1,1)$. We also analytically and graphically give the permissible range of macroscopic quantities for stability. For the athermal systems study the density $\rho$ can be any value, whereas the momentum $u$ should be within an area centered around the zero point. The exact shape of this region is determined by the choice of velocity discretization.

\newpage
\section*{Appendix: First order populations for the D2Q9 lattice with standard polynomial quasi-equilibria}
\label{app:d2q91st}
\begin{equation}
\begin{split}
\frac{\omega}{1-\omega}f^{1\ldots 5,(1)} =& \frac{4}{9}\left(-u_1\frac{\partial}{\partial x_1}\rho - u_2\frac{\partial}{\partial x_2}\rho + \left(1+\frac{3}{2}u^2\right)\frac{\partial}{\partial x_1}\rho u_1  \right. \\ & \qquad  \left.+ \left(1+\frac{3}{2}u^2\right)\frac{\partial}{\partial x_2}\rho u_2 - 3u_1\frac{\partial}{\partial x_1}\rho u_1^2 - 3u_2\frac{\partial}{\partial x_2}\rho u_2^2  \right. \\ & \qquad  \left.- 3u_1\frac{\partial}{\partial x_2}\rho u_1u_2 - 3u_2\frac{\partial}{\partial x_1}\rho u_1 u_2\right),
\\ & \frac{1}{9}\left( 2u_1\frac{\partial}{\partial x_1}\rho - u_2\frac{\partial}{\partial x_2}\rho + \left(-2-3u_1^2+\frac{3}{2}u_2^2\right)\frac{\partial}{\partial x_1}\rho u_1 \right. \\ & \qquad   + \left(1-3u_1^2+\frac{3}{2}u_2^2\right)\frac{\partial}{\partial x_2}\rho u_2+ 6u_1\frac{\partial}{\partial x_1}\rho u_1^2 \\ & \qquad + \frac{3}{2}\frac{\partial}{\partial x_1}\rho u_2^2   -3u_2\frac{\partial}{\partial x_2}\rho u_2^2 - 3u_2\frac{\partial}{\partial x_1}\rho u_1u_2   \\ & \qquad  \left.+ \left(3+6u_1\right)\frac{\partial}{\partial x_2}\rho u_1u_2 \right),
\\ & \frac{1}{9}\left( -u_1\frac{\partial}{\partial x_1}\rho + 2 u_2\frac{\partial}{\partial x_2}\rho + \left(1+\frac{3}{2}u_1^2-3u_2^2\right)\frac{\partial}{\partial x_1}\rho u_1 \right. \\ & \qquad   + \left(-2+\frac{3}{2}u_1^2-3u_2^2\right)\frac{\partial}{\partial x_2}\rho u_2 + 6u_2\frac{\partial}{\partial x_2}\rho u_2^2 \\ & \qquad  \frac{3}{2}\frac{\partial}{\partial x_2}\rho u_1^2  -3u_1\frac{\partial}{\partial x_1}\rho u_1^2 - 3u_1\frac{\partial}{\partial x_2}\rho u_1u_2 \\ & \qquad  \left. + \left(3+6u_2\right)\frac{\partial}{\partial x_1}\rho u_1u_2 \right),
\\ & \frac{1}{9}\left( 2u_1\frac{\partial}{\partial x_1}\rho - u_2\frac{\partial}{\partial x_2}\rho + \left(-2-3u_1^2+\frac{3}{2}u_2^2\right)\frac{\partial}{\partial x_1}\rho u_1 \right. \\ & \qquad  + \left(1-3u_1^2+\frac{3}{2}u_2^2\right)\frac{\partial}{\partial x_2}\rho u_2 + 6u_1\frac{\partial}{\partial x_1}\rho u_1^2\\ & \qquad   - \frac{3}{2}\frac{\partial}{\partial x_1}\rho u_2^2  -3u_2\frac{\partial}{\partial x_2}\rho u_2^2 - 3u_2\frac{\partial}{\partial x_1}\rho u_1u_2 \\ & \qquad \left. + \left(-3+6u_1\right)\frac{\partial}{\partial x_2}\rho u_1u_2 \right),
\\ & \frac{1}{9}\left( -u_1\frac{\partial}{\partial x_1}\rho + 2 u_2\frac{\partial}{\partial x_2}\rho + \left(1+\frac{3}{2}u_1^2-3u_2^2\right)\frac{\partial}{\partial x_1}\rho u_1 \right. \\ & \qquad  + \left(-2+\frac{3}{2}u_1^2-3u_2^2\right)\frac{\partial}{\partial x_2}\rho u_2+ 6u_2\frac{\partial}{\partial x_2}\rho u_2^2 \\ & \qquad - \frac{3}{2}\frac{\partial}{\partial x_2}\rho u_1^2  -3u_1\frac{\partial}{\partial x_1}\rho u_1^2 - 3u_1\frac{\partial}{\partial x_2}\rho u_1u_2 \\ & \qquad \left. + \left(-3+6u_2\right)\frac{\partial}{\partial x_15}\rho u_1u_2 \right).
\end{split}
\end{equation}
\begin{equation}
\begin{split}
\frac{\omega}{1-\omega}f^{6\ldots 8,(1)} = &  \frac{1}{36}\left( 2u_1\frac{\partial}{\partial x_1}\rho + 2u_2\frac{\partial}{\partial x_2}\rho + 3u_1\frac{\partial}{\partial x_2}\rho + 3u_2\frac{\partial}{\partial x_1}\rho \right.
\\ & \qquad  + \left(-2-3u_1^2-9u_1u_2-3u_2^2\right)\frac{\partial}{\partial x_1}\rho u_1 
\\ & \qquad + \left(-2-3u_1^2-9u_1u_2-3u_2^2\right)\frac{\partial}{\partial x_2}\rho u_2 
\\ & \qquad - 3\frac{\partial}{\partial x_1}\rho u_2 - 3\frac{\partial}{\partial x_2}\rho u_1 - 3\frac{\partial}{\partial x_1}\rho u_2^2 - 3\frac{\partial}{\partial x_2}\rho u_1^2 \\ & \qquad  + \left(6u_1+9u_2\right)\frac{\partial}{\partial x_1}\rho u_1^2 + \left(9u_1+6u_2\right)\frac{\partial}{\partial x_2}\rho u_2^2
\\ & \qquad  + \left(-6+9u_1+6u_2\right)\frac{\partial}{\partial x_1}\rho u_1u_2 
\\ & \qquad \left. + \left(-6+9u_1+6u_2\right)\frac{\partial}{\partial x_2}\rho u_1u_2\right),
\\ & \frac{1}{36}\left( 2u_1\frac{\partial}{\partial x_1}\rho + 2u_2\frac{\partial}{\partial x_2}\rho - 3u_1\frac{\partial}{\partial x_2}\rho - 3u_2\frac{\partial}{\partial x_1}\rho \right. \\ & \qquad + \left(-2-3u_1^2+9u_1u_2-3u_2^2\right)\frac{\partial}{\partial x_1}\rho u_1 
\\ & \qquad + \left(-2-3u_1^2+9u_1u_2-3u_2^2\right)\frac{\partial}{\partial x_2}\rho u_2 \\ & \qquad + 3\frac{\partial}{\partial x_1}\rho u_2 + 3\frac{\partial}{\partial x_2}\rho u_1 + 3\frac{\partial}{\partial x_1}\rho u_2^2 - 3\frac{\partial}{\partial x_2}\rho u_1^2 \\ & \qquad  + \left(6u_1-9u_2\right)\frac{\partial}{\partial x_1}\rho u_1^2 + \left(-9u_1+6u_2\right)\frac{\partial}{\partial x_2}\rho u_2^2
\\ & \qquad  + \left(-6-9u_1+6u_2\right)\frac{\partial}{\partial x_1}\rho u_1u_2 
\\ & \qquad \left. + \left(6-9u_1+6u_2\right)\frac{\partial}{\partial x_2}\rho u_1u_2\right),
\\ &\frac{1}{36}\left( 2u_1\frac{\partial}{\partial x_1}\rho + 2u_2\frac{\partial}{\partial x_2}\rho + 3u_1\frac{\partial}{\partial x_2}\rho + 3u_2\frac{\partial}{\partial x_1}\rho \right. \\ & \qquad + \left(-2-3u_1^2-9u_1u_2-3u_2^2\right)\frac{\partial}{\partial x_1}\rho u_1 
\\ & \qquad + \left(-2-3u_1^2-9u_1u_2-3u_2^2\right)\frac{\partial}{\partial x_2}\rho u_2 \\ & \qquad  - 3\frac{\partial}{\partial x_1}\rho u_2 - 3\frac{\partial}{\partial x_2}\rho u_1 + 3\frac{\partial}{\partial x_1}\rho u_2^2 + 3\frac{\partial}{\partial x_2}\rho u_1^2 \\ & \qquad  + \left(6u_1+9u_2\right)\frac{\partial}{\partial x_1}\rho u_1^2 + \left(9u_1+6u_2\right)\frac{\partial}{\partial x_2}\rho u_2^2 \\ & \qquad  + \left(6+9u_1+6u_2\right)\frac{\partial}{\partial x_1}\rho u_1u_2 
\\ & \qquad \left. + \left(6+9u_1+6u_2\right)\frac{\partial}{\partial x_2}\rho u_1u_2\right),
\end{split}
\end{equation}
\begin{equation}
\begin{split}
\frac{\omega}{1-\omega}f^{ 9,(1)} = & \frac{1}{36}\left( 2u_1\frac{\partial}{\partial x_1}\rho + 2u_2\frac{\partial}{\partial x_2}\rho - 3u_1\frac{\partial}{\partial x_2}\rho - 3u_2\frac{\partial}{\partial x_1}\rho \right. \\ & \qquad + \left(-2-3u_1^2+9u_1u_2-3u_2^2\right)\frac{\partial}{\partial x_1}\rho u_1 
\\ & \qquad + \left(-2-3u_1^2+9u_1u_2-3u_2^2\right)\frac{\partial}{\partial x_2}\rho u_2 \\ & \qquad  + 3\frac{\partial}{\partial x_1}\rho u_2 + 3\frac{\partial}{\partial x_2}\rho u_1 - 3\frac{\partial}{\partial x_1}\rho u_2^2 + 3\frac{\partial}{\partial x_2}\rho u_1^2 \\ & \qquad  + \left(6u_1-9u_2\right)\frac{\partial}{\partial x_1}\rho u_1^2 + \left(-9u_1+6u_2\right)\frac{\partial}{\partial x_2}\rho u_2^2 \\ & \qquad  + \left(6-9u_1+6u_2\right)\frac{\partial}{\partial x_1}\rho u_1u_2 
\\ & \qquad \left. + \left(-6-9u_1+6u_2\right)\frac{\partial}{\partial x_2}\rho u_1u_2\right)
\end{split}
\end{equation}

\end{document}